\documentstyle[graphicx,pifont,amssymb,amsmath]{mn}

\title[{\it Chandra\/} Observations of NGC\,3256]
{{\it Chandra\/} Observations of the Luminous IR Galaxy NGC\,3256}

\author[Lira etal.]
{P.~Lira,$^{1,2}$  M.~Ward,$^1$ A.~Zezas,$^3$ A.~Alonso-Herrero$^{4,5}$ S.~Ueno,$^6$\\
$^1$ Department of Physics \& Astronomy, University of Leicester, Leicester LE1 7RH, UK\\
$^2$ Departamento de Astronom\'{\i}a, Universidad de Chile, Casilla 36-D, Santiago, Chile\\
$^3$ Harvard-Smithsonian Center for Astrophysics, 60 Garden St., Cambridge MA, 02138, USA\\ 
$^4$ Department of Physical Science, University of Hertfordshire, Hatfield, Herts AL10 9AB, UK\\
$^5$ Steward Observatory, University of Arizona, Tucson, AZ 85721, USA\\
$^6$ Space Utilization Research Program, National Space Development Agency of Japan (NASDA), Tsukuba Space Center,\\ 
2-1-1 Sengen, Tsukuba, Ibaraki 305, Japan\\}

\begin{document}

\maketitle

\begin{abstract} 

We present a detailed analysis of high-resolution {\it Chandra\/}
observations of the merger system NGC3256, the most IR luminous galaxy
in the nearby universe. The X-ray data show that several discrete
sources embedded in complex diffuse emission contribute $\ga 20\%$ of
the total emission ($L_{\rm x}^{\rm tot} \sim 8 \times 10^{41}$ ergs
s$^{-1}$ in the 0.5-10 keV energy range). The compact sources are hard
and extremely bright and their emission is probably dominated by
accretion driven processes. Both galaxy nuclei are detected with
$L_{\rm x} \sim 3-10 \times 10^{40}$ ergs s$^{-1}$. No evidence is
found for the presence of an active nucleus in the southern nucleus,
contrary to previous speculation. Once the discrete sources are
removed, the diffuse component has a soft spectrum which can be
modelled by the superposition of 3 thermal plasma components with
temperatures $kT = 0.6, 0.9$ and 3.9 keV. Alternatively, the latter
component can be described as a power-law with index $\Gamma \sim
3$. Some evidence is found for a radial gradient of the amount of
absorption and temperature of the diffuse component. We compare the
X-ray emission with optical, H$\alpha$ and NICMOS images of NGC3256
and find a good correlation between the inferred optical/near-IR and
X-ray extinctions. Although Inverse Compton scattering could be
important in explaining the hard X-rays seen in the compact sources
associated with the nuclei, the observed diffuse emission is probably
of thermal origin. The observed X-ray characteristics support a
scenario in which the powerful X-ray emission is driven solely by the
current episode of star formation.

\end{abstract}


\section{Introduction}

\begin{figure*} 
\centering
\caption{{\it Chandra\/} images of NGC3256 in the 0.3-1.5 (left) 
and 1.5-10 (right) keV energy ranges. The images extend $\sim 60$
arcsecs on each side. North is at the top and East in on the
left. Discrete sources are labelled by increasing order of right
ascension. Sources 1 and 14 are not included in the figure (see Table
1 for celestial coordinates).}
\end{figure*}

The IRAS all-sky survey revealed a class of galaxies with infrared
luminosities which overlap the bolometric luminosity of quasars. The
vast amounts of energy released by these Luminous Infra-Red Galaxies
(LIRGs: $L_{\rm IR} > 10^{11}\ L_{\sun}$) and Ultra Luminous Infra-Red
Galaxies (ULIRGs: $L_{\rm IR} > 10^{12}\ L_{\sun}$) opened a question
that has still not been completely resolved: what is the primary
source of their power?  The two favored answers are extreme starburst
activity or the presence of an active nucleus.  Important clues came
from the observed correlation between IR luminosity and the presence
of an AGN, showing that the fraction of Seyfert nuclei detected in
optical surveys increases from $\sim 5\%$ to $\sim 50\%$ for objects
with $\log L_{\rm IR} \sim 10.5\ L_{\sun}$ and $\log L_{\rm IR} \sim
12.5\ L_{\sun}$ respectively (Sanders \& Mirabel 1996).

The definite presence (or absence) of an AGN in infrared luminous
systems is difficult to test, since tens or even hundreds of
magnitudes of visual extinction are routinely derived for the central
regions of these galaxies.  However, recent results from mid-IR
observations using {\it ISO\/}, which can probe regions with
extinctions as high as $A_{V} \sim 50$ magnitudes, show that $70-80\%$
percent of ULIRGS are predominantly powered by star formation activity
(Genzel etal 1998; Lutz, Veilleux \& Genzel 1999).

Since nearly all ULIRGS have distorted morphologies it has been
suggested that galactic interactions displace large masses of
molecular gas into the central few kpc of these systems, fueling a
powerful starburst or an accretion powered AGN. Several possible
evolutionary scenarios have been proposed to link ULIRGs with QSOs,
`super-starbursts' and elliptical galaxies (Sanders etal 1988; Joseph
\& Wright 1985; see also L\'{\i}pari etal 2001 and references
therein).

NGC3256 is a spectacular merger system. Its IR luminosity of $6
\times 10^{11}\ L_{\sun}$ makes it the most luminous object in the
local ($z<0.01$) universe \footnote{A distance of 56 Mpc ($z=0.009$,
H$_\circ=50$ km s$^{-1}$ Mpc$^{-1}$) has been used throughout this
paper.}. Evidence of a recent interaction is the presence of 200 kpc
extended tails and a highly chaotic nuclear region (Graham etal
1984). Direct evidence for the presence of a powerful starburst in the
galaxy comes from the detection of strong absorption features from
massive stars (Doyon, Joseph \& Wright 1994, L\'{\i}pari etal 2000).
The central region of the starburst ($\sim 5$ kpc across) is
characterized by strong near-IR line emission, high metallicities, and
outflows. (Moorwood \& Oliva 1994; L\'{\i}pari etal 2000). A double
nucleus has been detected at radio and near-IR frequencies (Norris \&
Forbes 1995; Kotilainen etal 1996), suggesting that NGC3256 is not a
very advanced merger system. While the northern nucleus has always
been recognised as a pure starburst, the presence of a hidden AGN in
the southern nucleus has remained a matter of speculation.

X-ray observations of NGC3256 were obtained previously with {\it
ROSAT\/} and {\it ASCA\/}. From the analysis of {\it ASCA\/} data
Moran, Lehnert \& Helfand (1999, hereafter MLH) determined a total
luminosity in the 0.5-10 keV energy range of $1.6 \times 10^{42}$ ergs
s$^{-1}$. If powered solely by star formation, NGC3256 would represent
the top end of the X-ray luminosity distribution of starburst
galaxies. MLH also argued that the observed power law tail could not
be explained by the presence of an AGN or a population of
Galactic-type X-ray binaries and suggested that Inverse Compton
scattering could provide a substantial fraction of the observed high
energy component.

In order to understand the nature of the NGC3256 powerful emission,
observations at high spatial resolution are required. High resolution
X-ray observations are particularly suited for this aim, since hard
X-rays are able to probe deep into the most obscured regions and give
vital information about hot gaseous components commonly associated
with strong starbursts and which are undetectable at other
wavelengths.  In this paper we report recent high resolution
observations of NGC3256 made using the {\it Chandra\/} X-ray
Observatory. This paper is organised as follows: section 2 gives
details of the data analysis procedure while sections 3 and 4 focus on
the characteristics of the population of compact sources and the
diffuse emission respectively; in section 5 a brief comparison between
the {\it Chandra\/} and {\it ASCA\/} observations is given; a full
discussion is presented in section 6 and conclusions and summary can
be found in section 7.

\section{Observations and data analysis}

\begin{table}
\caption{Detected discrete sources in NGC3256. Background subtracted
source counts ($SC$) above 0.3 keV are given for each source. The
significance of the detections was determined as $SC/\sigma_{B}$,
where $\sigma_{B}$ is the standard deviation of the background and was
computed as $1 + \sqrt{\rm{Background\ counts} + 0.75}$. This correction
to the standard deviation gives a more appropriate estimate of
Poissonian errors for cases of low number counts (Gehrels 1986).}
\centering
\begin{tabular}{lcrr} \hline
Source & Position (J2000) & Source Counts & Significance \\ \hline
1       &10 27 48.02	 -43 54 26.5 &22.8  $\pm$  6.2   &25.34 \\
2       &10 27 50.04	 -43 54 20.5 &50.6  $\pm$  10.9  &18.58 \\
3       &10 27 50.74	 -43 54 16.2 &38.7  $\pm$  12.8  &11.53 \\
4       &10 27 50.93	 -43 53 43.4 &25.6  $\pm$  7.0   &19.15 \\
5       &10 27 51.23	 -43 53 58.8 &59.7  $\pm$  11.0  &24.00 \\
6       &10 27 51.23	 -43 54 10.9 &92.4  $\pm$  17.3  &18.27 \\
7(N)    &10 27 51.25	 -43 54 14.2 &712.2 $\pm$  30.6  &141.38 \\
8(S)    &10 27 51.22	 -43 54 19.4 &34.3  $\pm$  13.0  &10.09 \\
9       &10 27 51.64	 -43 54 09.7 &171.2 $\pm$  18.9  &36.27 \\
10      &10 27 51.76	 -43 54 13.6 &261.6 $\pm$  20.7  &60.05  \\
11      &10 27 52.03	 -43 54 12.6 &119.8 $\pm$  15.1  &38.84 \\
12      &10 27 52.56	 -43 53 49.9 &48.3  $\pm$  8.5   &42.84 \\
13      &10 27 52.88	 -43 54 11.8 &26.7  $\pm$  7.9   &14.51 \\
14      &10 27 55.14	 -43 54 47.1 &88.5  $\pm$  10.7  &86.00 \\
\hline
\end{tabular}
\end{table}

A {\it Chandra\/} 28 ks observation of NGC3256 was obtained on the
5$^{th}$ of January 2000. For an assumed distance of 56 Mpc, 1 arcsec
(encircling $\sim 90\%$ of the photons from a point source at 1.5 keV
\footnote{see the {\it Chandra\/} Proposers' Observatory Guide v3.0,
December 2000, http://asc.harvard.edu/udocs/docs/POG/MPOG/})
corresponds to a linear size of 270 pc. The data were analysed using a
combination of {\it Chandra\/} X-ray Center (CXC) Ciao (V1) and
HERSARC XSPEC (V10.0) software. The galaxy was imaged using the
(back-illuminated) ACIS S3 chip, which is known to suffer from flares
of high background radiation. An examination of our observations,
after discarding all events outside the 0.3-10.0 keV energy range,
showed that in fact about half of the observing time was affected by
high background. The strategy adopted to ensure minimum contamination
from the high background episodes is described in sections 3 and
4.1. The average count rate during the observation was found to be
2.55 counts per second in the S3 chip, compared with the quiescent
0.79 counts per second measured in-flight \footnote{see the {\it
Chandra\/} Proposers' Observatory Guide v3.0, December 2000,
http://asc.harvard.edu/udocs/docs/POG/MPOG/}.  The three known bad
columns of the S3 chip were located far away ($\sim 5$ arcmin) from
our target and therefore did not induce any problems during the data
analysis.

\begin{figure*}
\centering
\includegraphics[angle=270,scale=0.28]{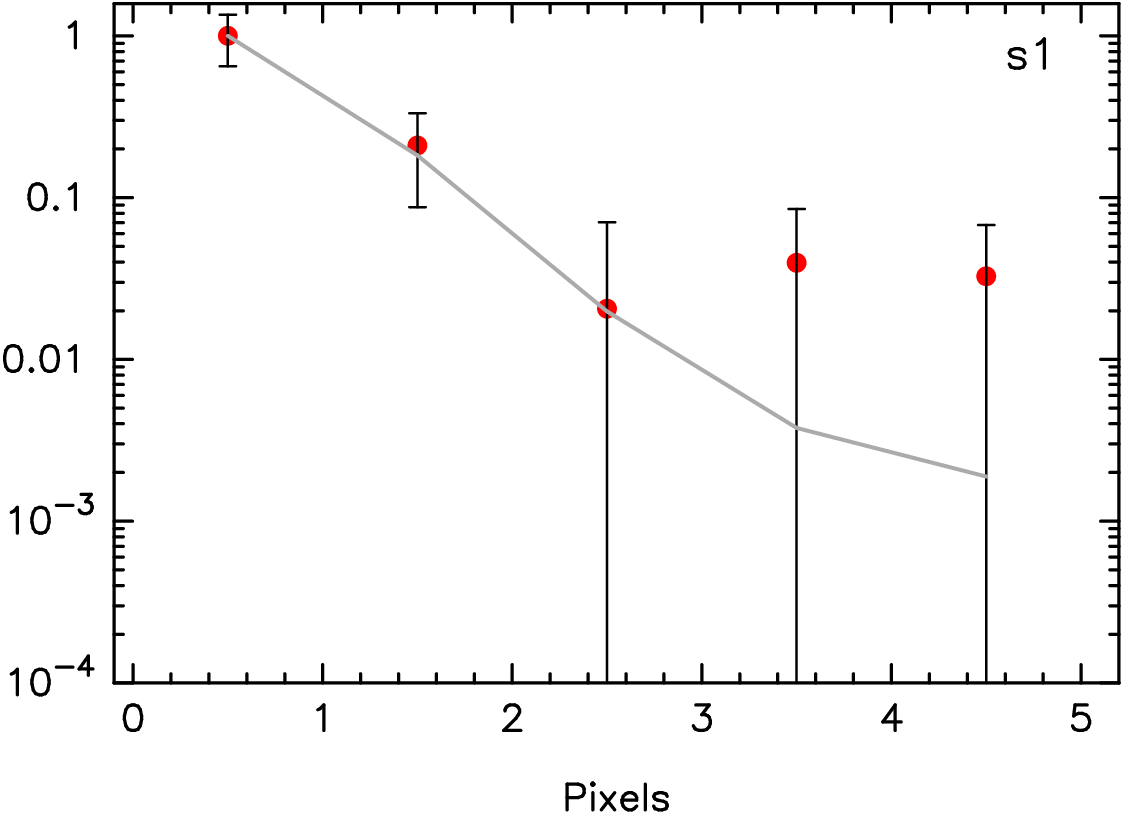}\hspace{0.3cm}
\includegraphics[angle=270,scale=0.28]{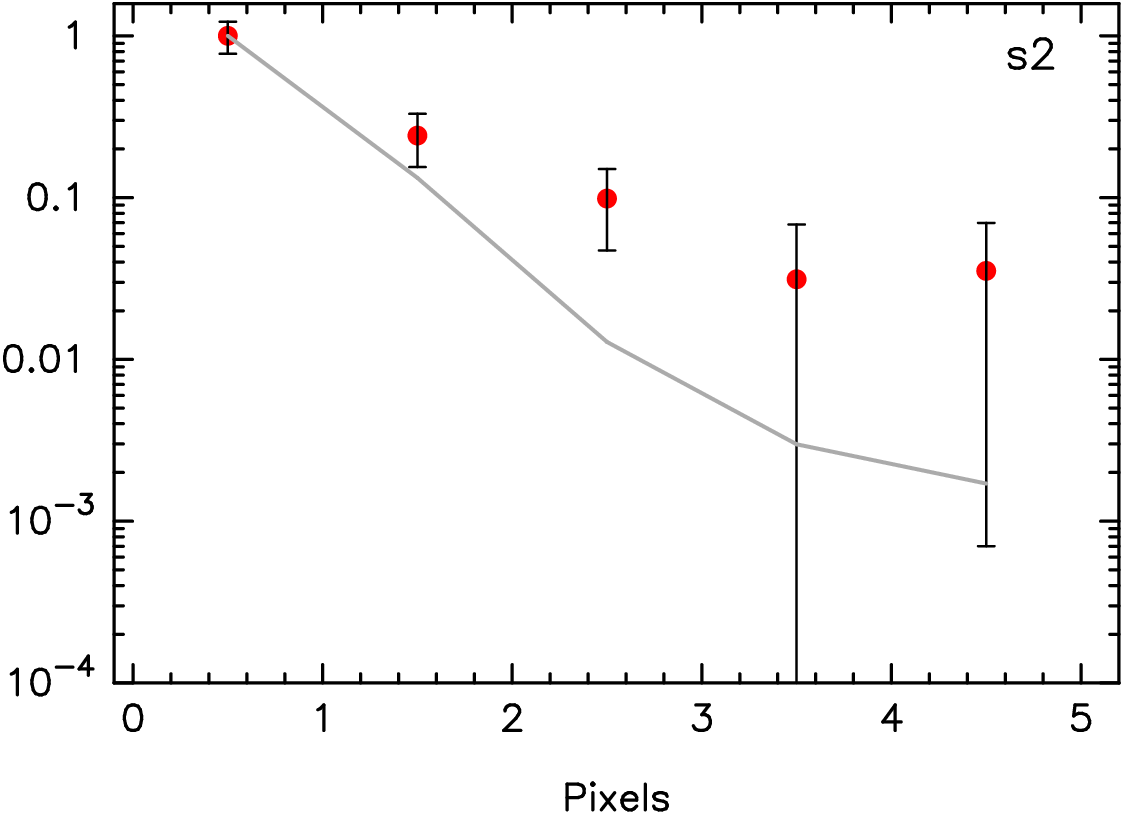}\hspace{0.3cm}
\includegraphics[angle=270,scale=0.28]{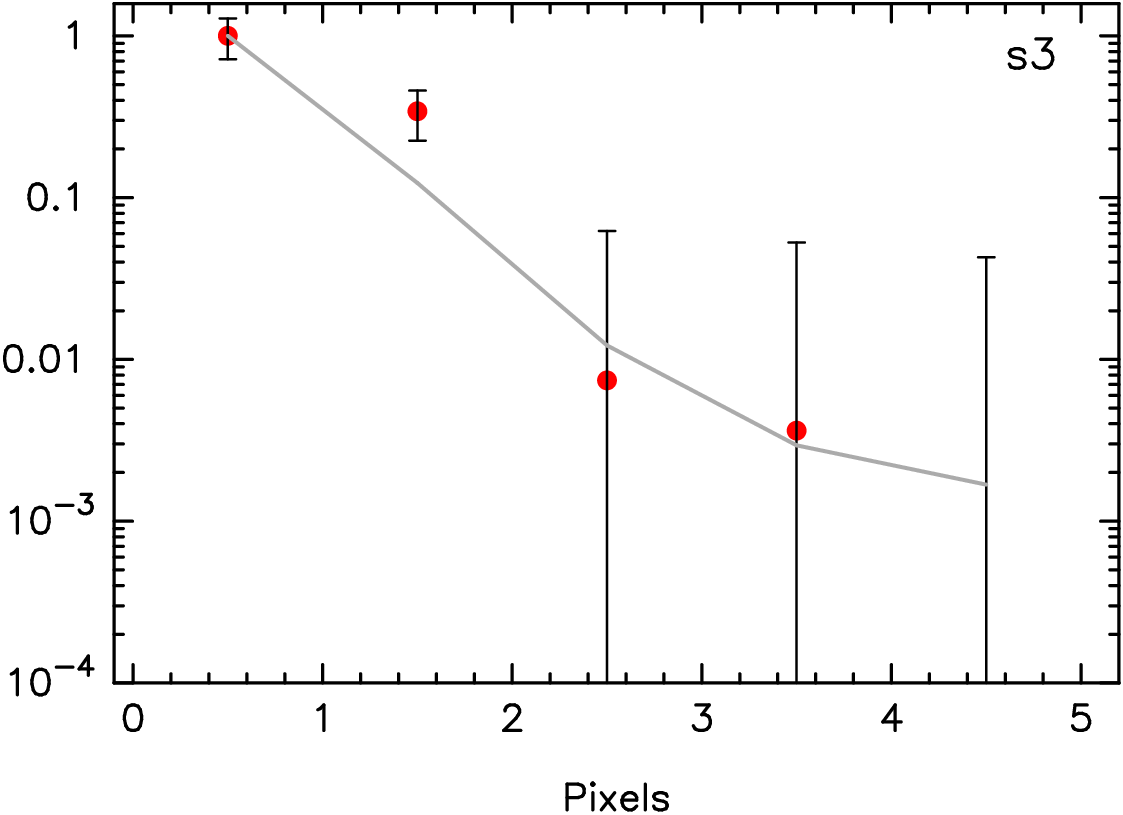}\vspace{0.4cm}
\includegraphics[angle=270,scale=0.28]{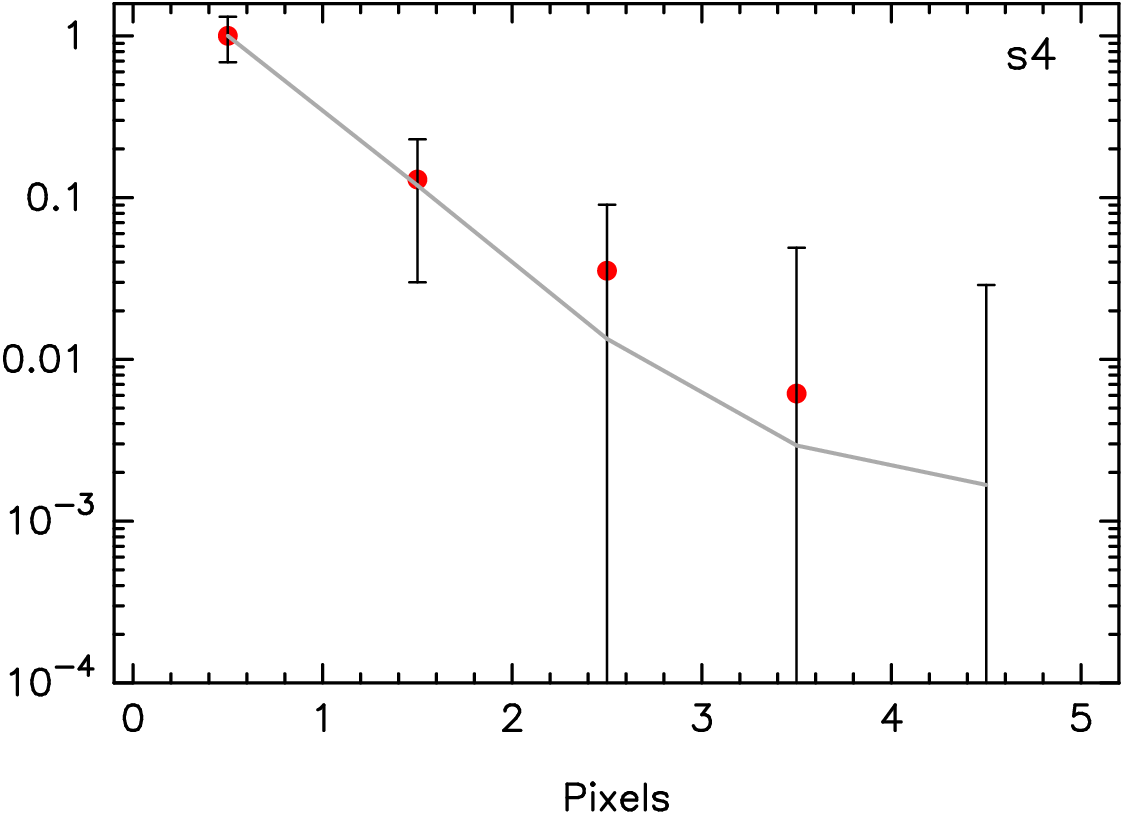}\hspace{0.3cm}
\includegraphics[angle=270,scale=0.28]{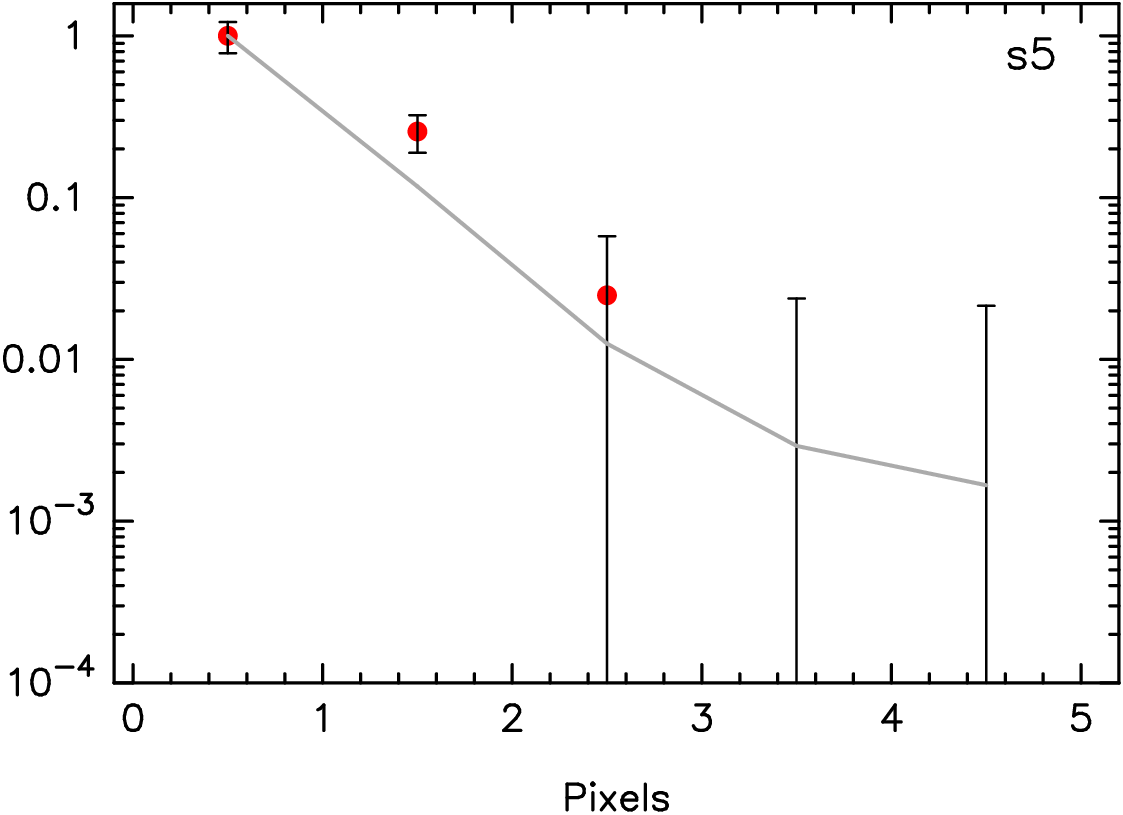}\hspace{0.3cm}
\includegraphics[angle=270,scale=0.28]{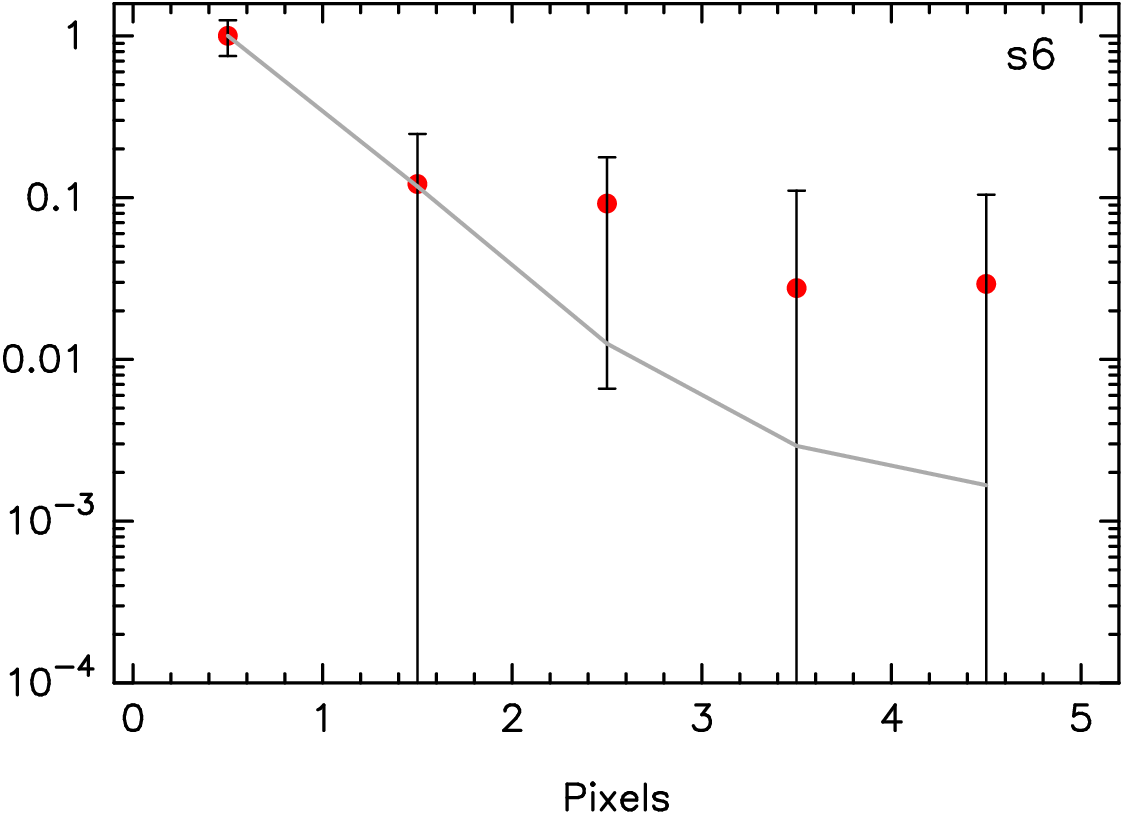}\vspace{0.4cm}
\includegraphics[angle=270,scale=0.28]{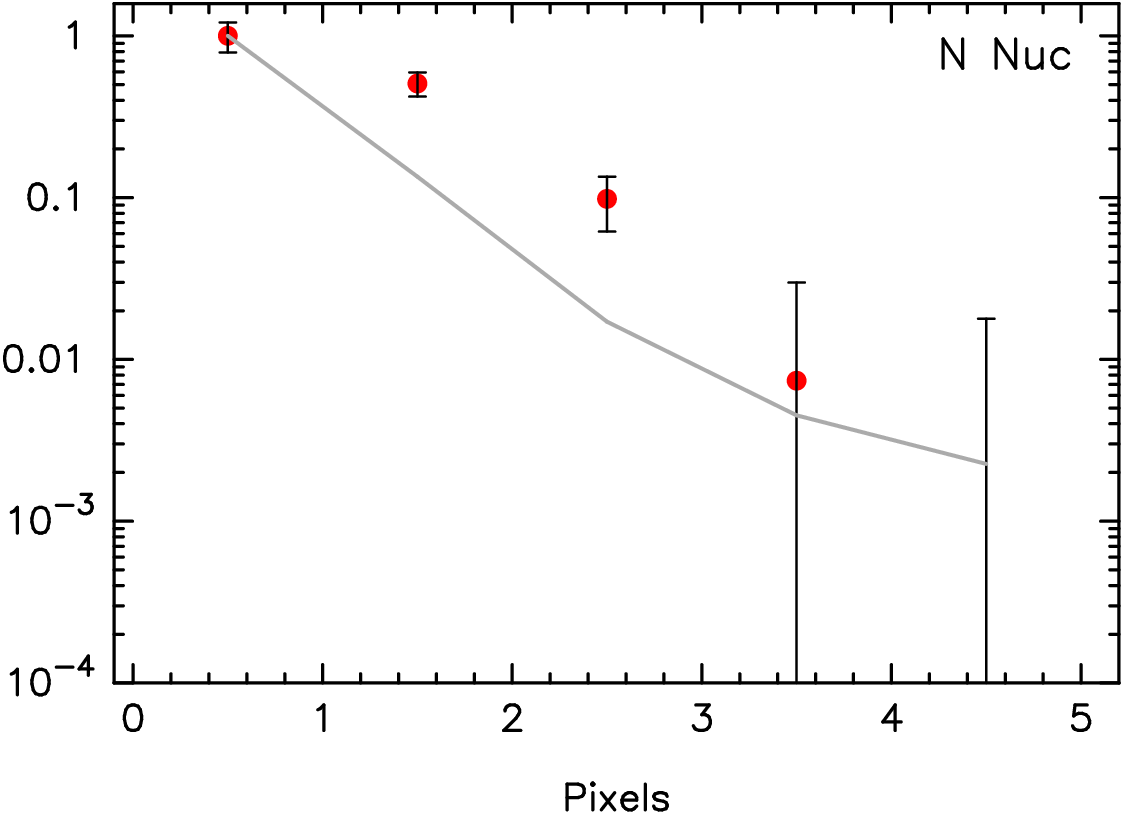}\hspace{0.3cm}
\includegraphics[angle=270,scale=0.28]{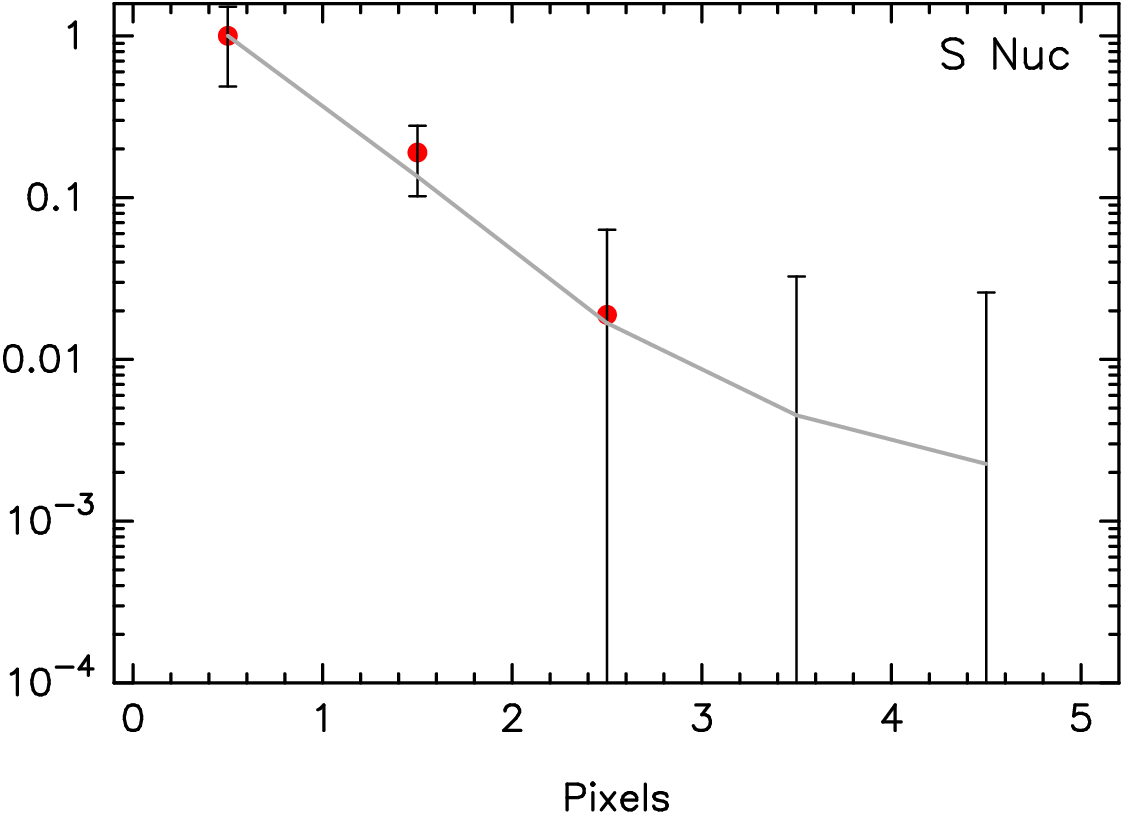}\hspace{0.3cm}
\includegraphics[angle=270,scale=0.28]{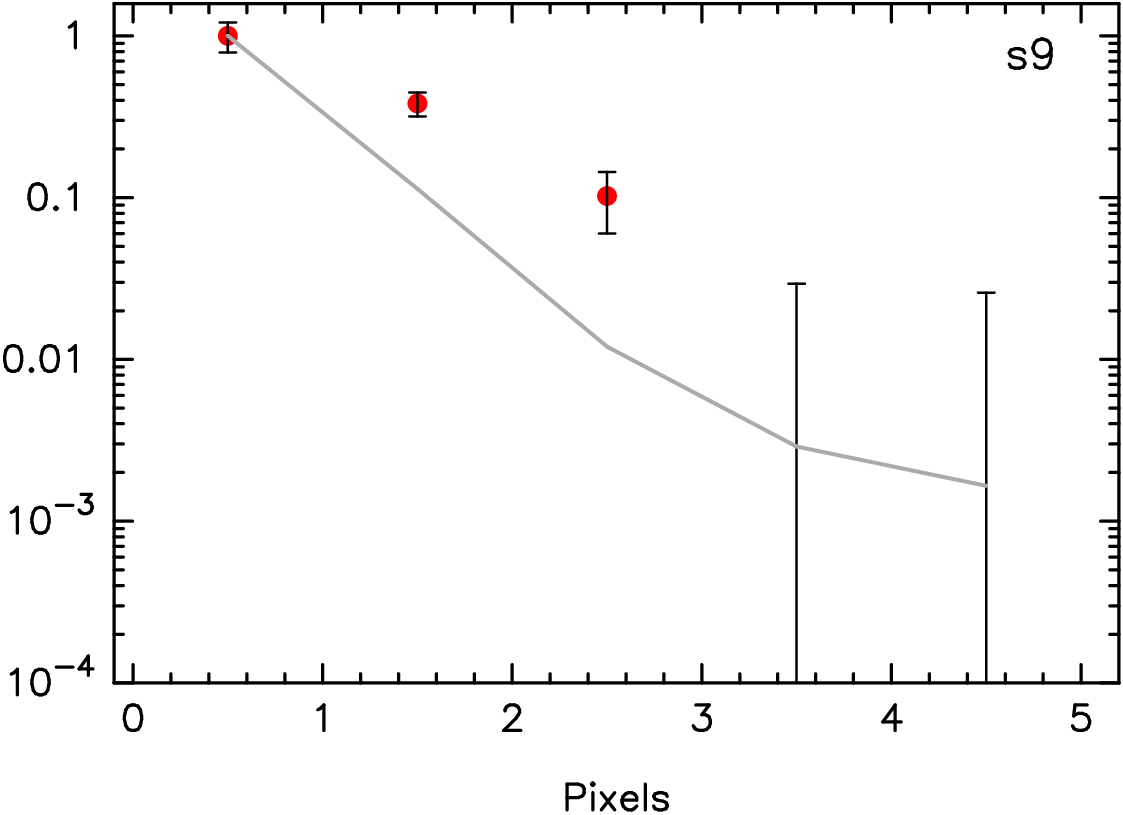}\vspace{0.4cm}
\includegraphics[angle=270,scale=0.28]{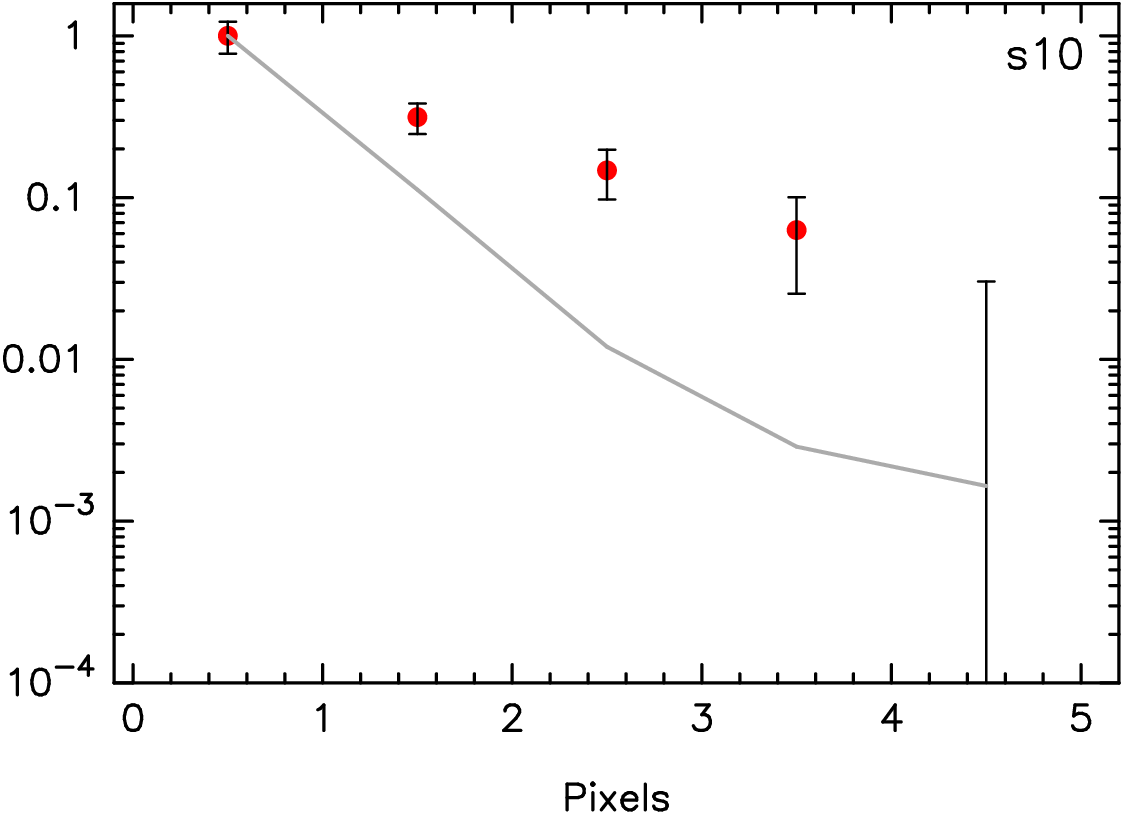}\hspace{0.3cm}
\includegraphics[angle=270,scale=0.28]{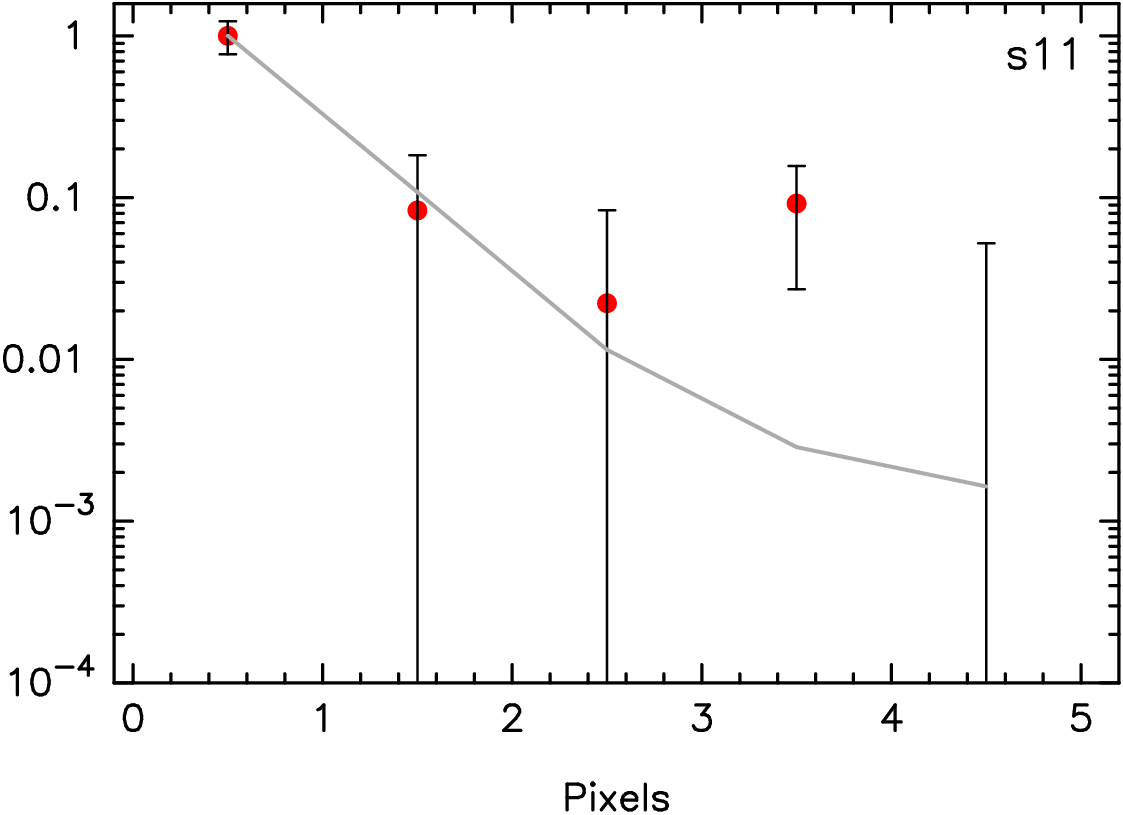}\hspace{0.3cm}
\includegraphics[angle=270,scale=0.28]{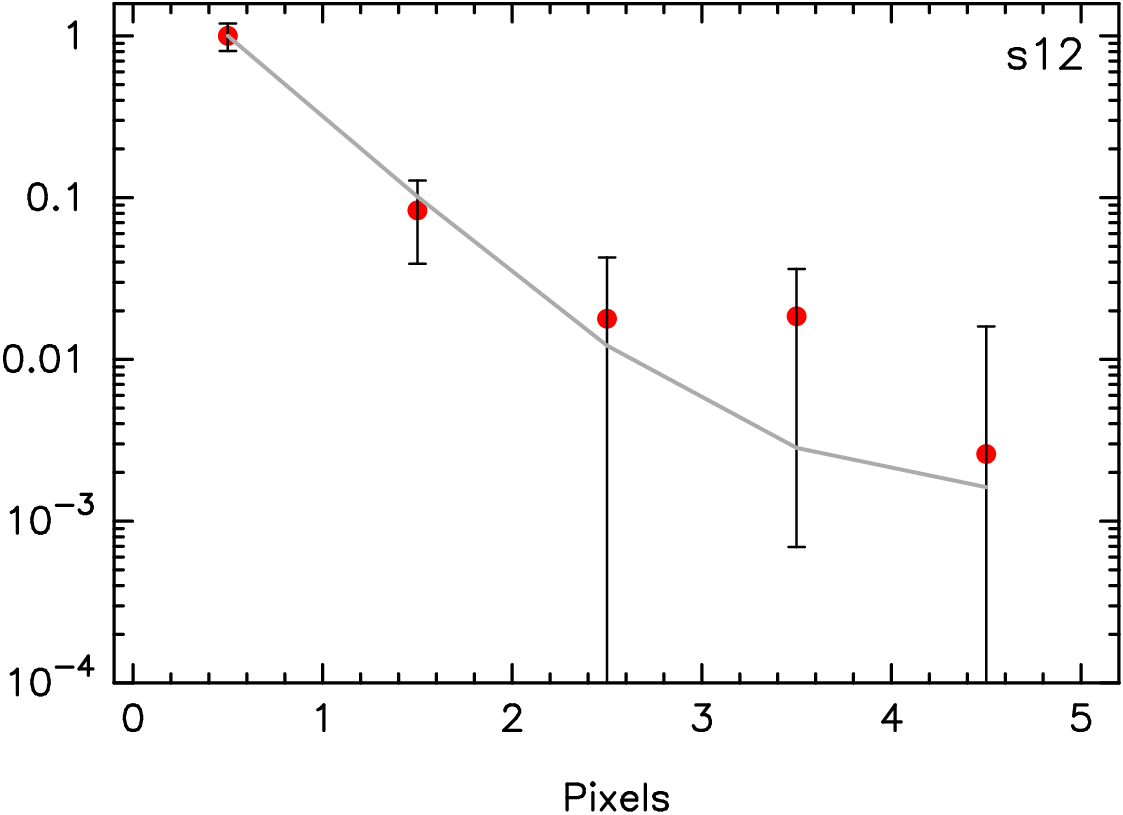}\vspace{0.4cm}
\includegraphics[angle=270,scale=0.28]{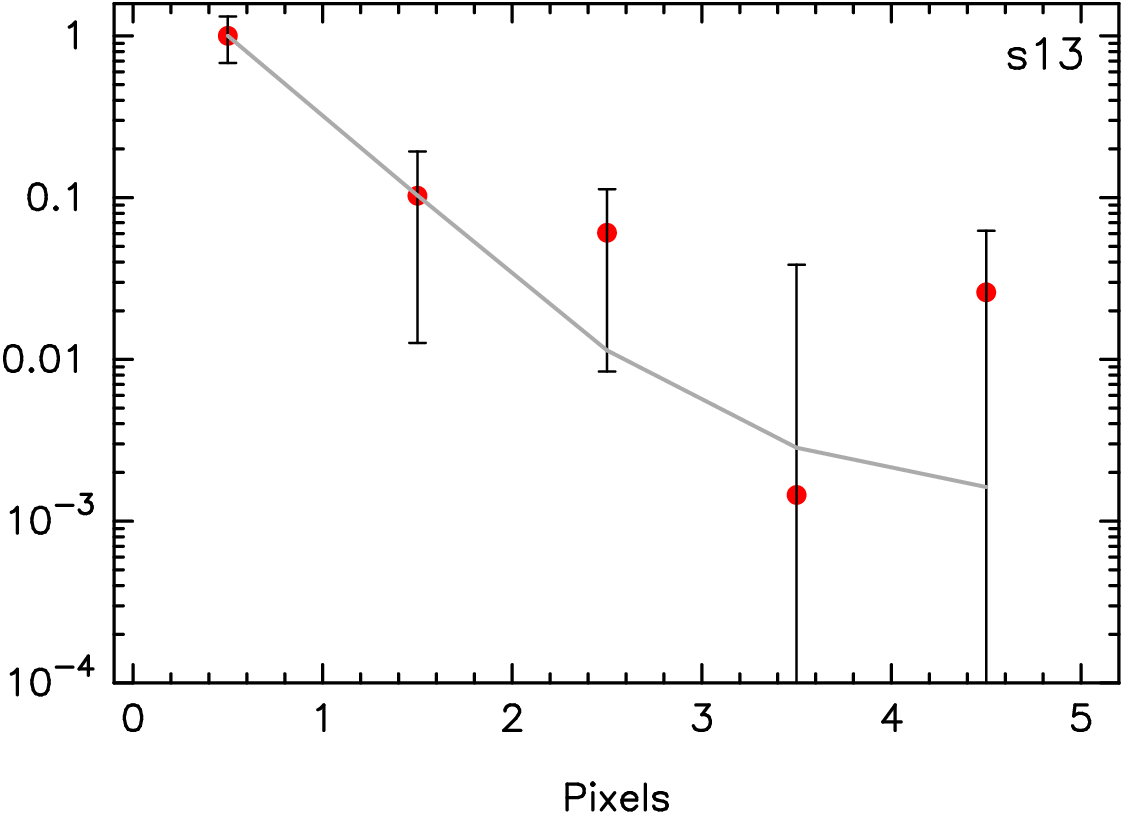}\hspace{0.3cm}
\includegraphics[angle=270,scale=0.28]{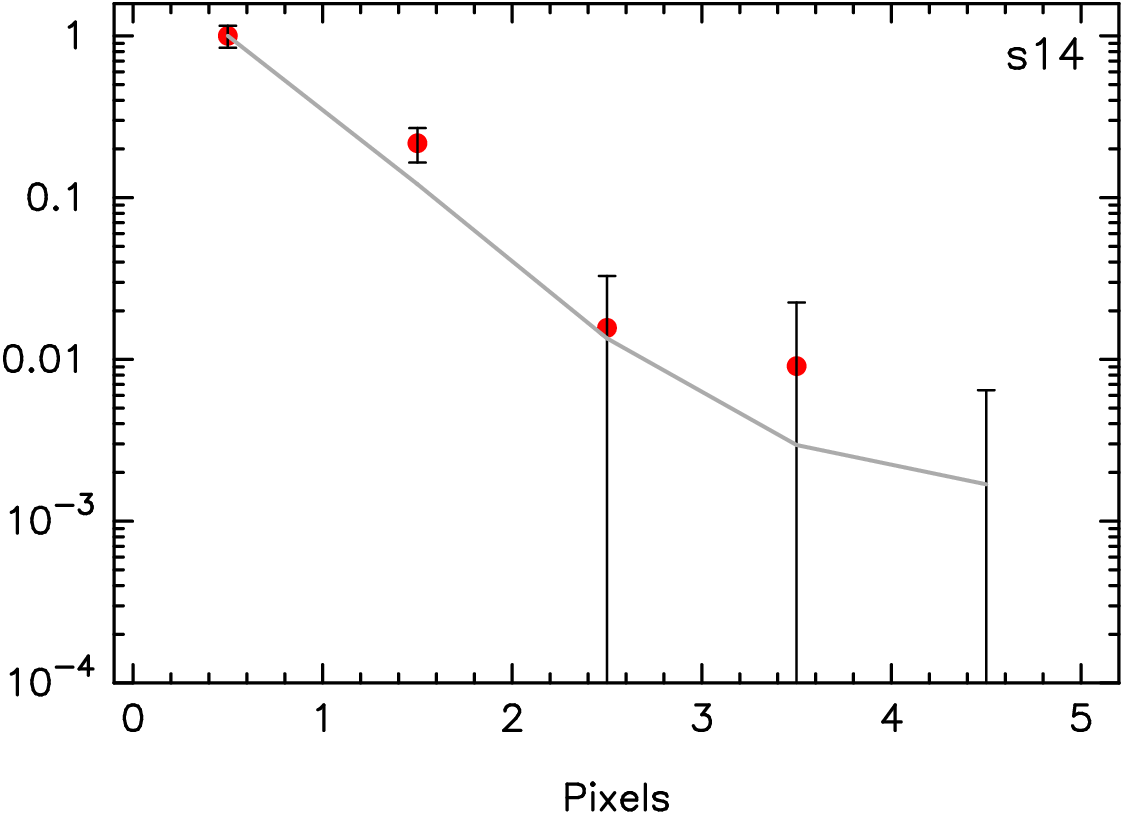}
\caption{Spatial profiles of the discrete sources obtained from a full
0.3-10 keV band image and compared with the model PSF (continuous
lines). The error bars correspond to 1$\sigma$ confidence levels. The
profiles for both nuclei (N and S) were obtained from a harder, 2-5
keV image.}
\end{figure*}

Simple aperture photometry shows that all the galactic emission is
contained within a circle of radius $\sim 60$ arcsec, while circles of
radius $\sim 27$ and $\sim 18$ arcsec encircle $\sim 95\%$ and $\sim
90\%$ of the counts, respectively. The large first radius is necessary
to contain the two most distant discrete sources (sources 1 and 14 in
Figure 1 and Table 1), which are located at $\sim 40 - 55$ arcsec from
the central peak of emission.

\begin{table*} 
\centering 
\caption{Fitted models for the emission from sources 7 (the
northern nucleus), 9 and 10. Hydrogen columns are given in units of
$10^{21}$ cm$^{-2}$ and plasma temperatures in units of keV. The
errors correspond to 90\% CL for 1 parameter of interest. {\it
Observed\/} fluxes are in units of $10^{-14}$ ergs s$^{-1}$ cm$^{-2}$
and {\it intrinsic\/} luminosities are in units of $10^{40}$ ergs
s$^{-1}$ (for an assumed distance of 56 Mpc).}
\begin{tabular}{lcccccccccc} \hline
Source   & Model& $N_{H}$            & $kT$ & $N_{H}$ & $\Gamma$ & $\chi^{2}_{\rm red}$(dof) & $F_{\rm x}$ & $L_{\rm x}$& $F_{\rm x}$ & $L_{\rm x}$ \\
         &      &                    &      &         &                  &                       & 0.5-2.5     & 0.5-2.5    & 2.5-10      & 2.5-10  \\
         &      &                    &      &         &                  &                       & (keV)       & (keV)      & (keV)       & (keV) \\ \hline
7(N) $a$ & PL   & - 		     &        -               &$3.8^{+0.8}_{-0.6}$  &$2.81^{+0.35}_{-0.28}$ & 1.81(32) & 6.77 & 6.88 & 3.90 & 1.52 \\[3pt]
7(N) $b$ & M+PL & $3.4^{+1.0}_{-0.9}$& $0.66^{+0.12}_{-0.10}$ &        $\dag$       &$2.25^{+0.38}_{-0.24}$ & 0.91(30) & 6.64 & 5.68 & 4.80 & 1.85 \\[3pt]
7(N) $c$ & M+PL & $Gal$              & $0.69^{+0.10}_{-0.06}$ & $5.8^{+8.0}_{-2.3}$ &$2.57^{+0.81}_{-0.33}$ & 0.84(30) & 7.01 & 8.50 & 4.53 & 1.79 \\[3pt]
9        & PL   & - 		     &        -               &$1.6^{+2.1}_{Gal}$   &$2.43^{+1.09}_{-0.80}$ & 1.04(10) & 1.59 & 0.93 & 0.91 & 0.35 \\[3pt]
10       & PL   & - 		     &        -               &$2.9^{+2.0}_{-1.7}$  &$2.91^{+1.20}_{-0.81}$ & 1.51(13) & 2.38 & 2.08 & 0.73 & 0.28 \\[3pt]
10       & M+PL & $Gal$              & $1.00^{+0.18}_{-0.23}$ &        $\dag$       &$1.67^{+0.40}_{-0.37}$ & 0.93(12) & 2.55 & 1.15 & 1.28 & 0.44 \\
\multicolumn{11}{l}{}\\[3pt]
\multicolumn{11}{l}{$\dag$: Same hydrogen column as applied to previous spectral component.}\\[3pt]
\multicolumn{11}{l}{Throughout this paper, whenever the column of gas fitted to a low temperature thermal component (probably associated}\\
\multicolumn{11}{l}{to a large-scale extended hot phase) presented an unrealistically low value of $N_{H}$, the Galactic hydrogen column ($N_{Gal} \sim$}\\
\multicolumn{11}{l}{$9\times10^{20}$ cm$^{-2}$, labelled as $Gal$) was adopted instead.}\\
\hline
\end{tabular}
\end{table*}

Due to a calibration error in early {\it Chandra\/} data, the
astrometry of the original observations of NGC3256 were affected by a
$\sim 8$ arcsec shift in absolute astrometry. After reprocessing of
the data at the CXC, which corrected for this problem, we found that
the position of the brightest X-ray compact source seen in the galaxy
(source 7 in Figure 1 and Table 1) was offset with respect to the
radio position for the northern nucleus (which is coincident, within 1
arcsec, with the position of the nucleus seen at optical and IR
wavelengths - Norris \& Forbes 1995) by only 0.44 arcsecs. Also, good
positional agreement is found between several knots of emission seen
in X-rays and optical wavelengths (see sections 6.2 and 6.3.1).

Figure 1 shows a $\sim 60 \times 60$ arcsec {\it Chandra\/} image of
NGC3256 in the 0.3-1.5 and 1.5-10 keV energy ranges. The soft image
shows a significant and spatially complex diffuse component. Several of
the discrete sources are more easily seen in the harder band, where the
diffuse emission has a much lower surface brightness. The emission is
dominated by a strong compact source located at the center of the X-ray
emitting region and labeled as source 7. As mentioned above, this
source has been identified with the northern nucleus of the galaxy.
Another, harder source is found 5 arcseconds to the south of this
position in the 1.5-10 keV band (source 8), and is identified as the
X-ray counterpart of the southern nucleus. The diffuse emission shows
complex morphology, with a strong central `Bar' running in a ES to NW
direction just above of the northern nucleus. Another interesting
feature is a southern `Arm' connecting sources 8 and 2, which leaves a
void of soft X-ray emission at the position of the southern nucleus.

\section{Discrete sources}

\begin{figure}
\centering
\includegraphics[angle=270,scale=0.3]{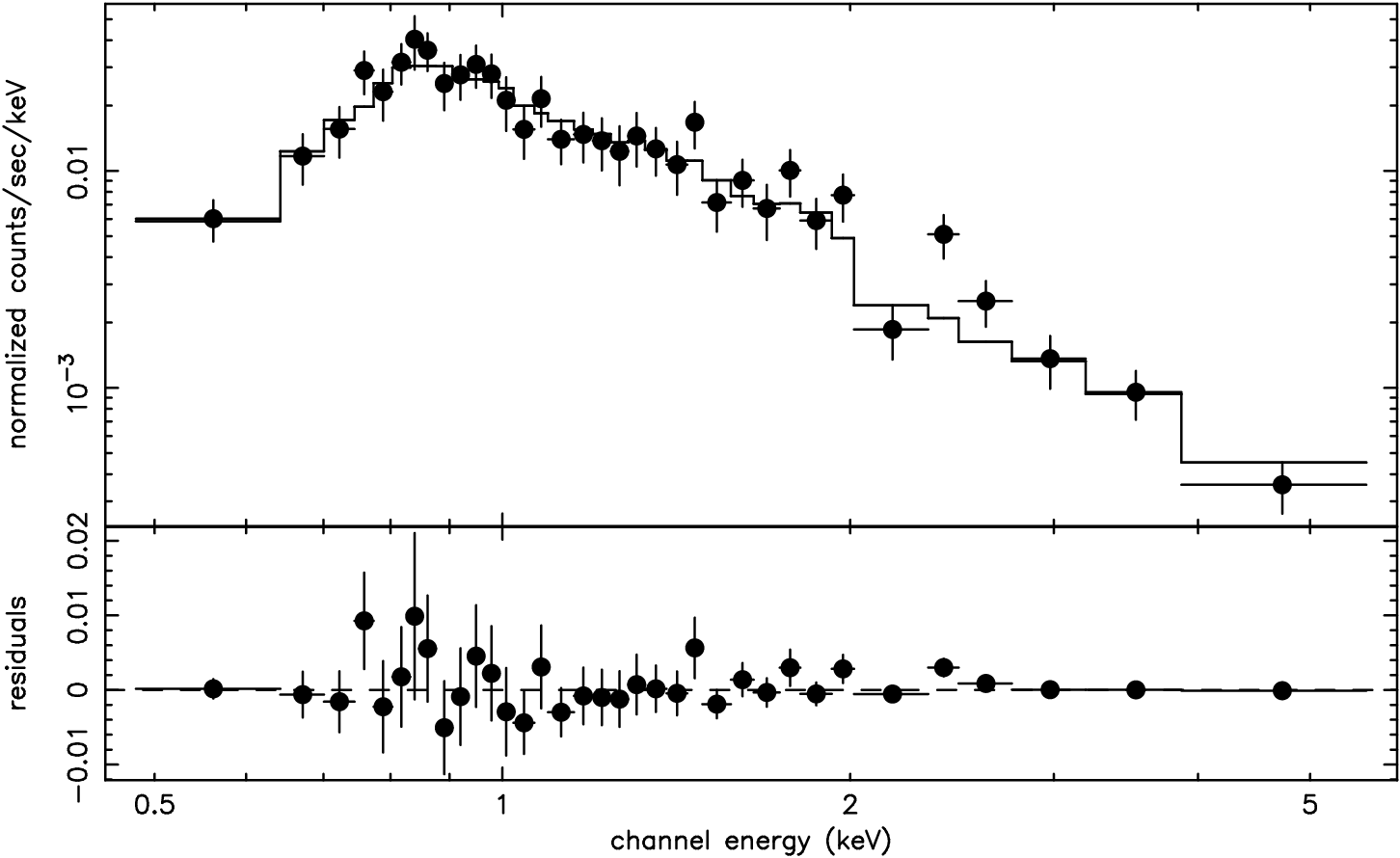}\\ \vspace{0.5cm}
\includegraphics[angle=270,scale=0.3]{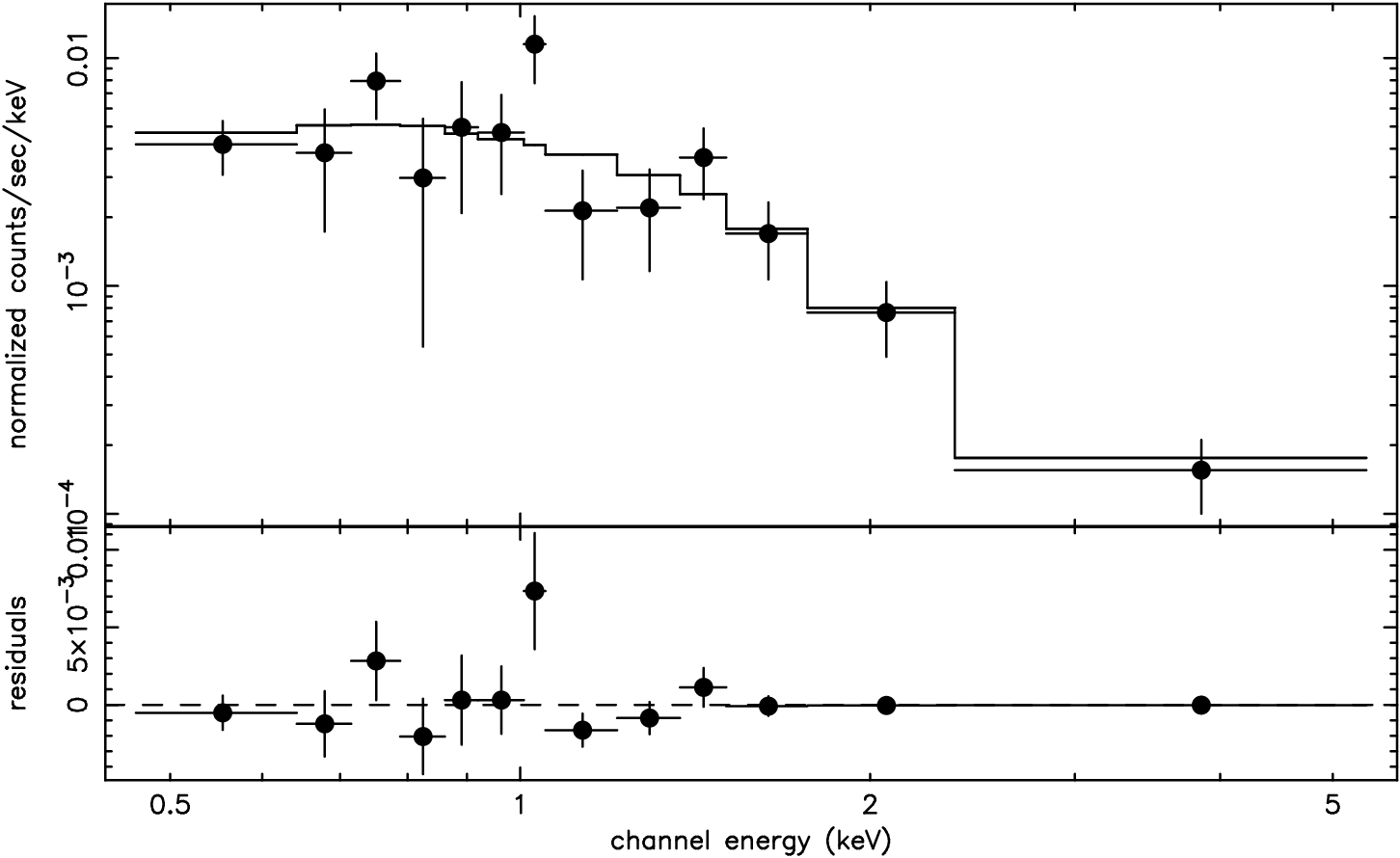}\\ \vspace{0.5cm}
\includegraphics[angle=270,scale=0.3]{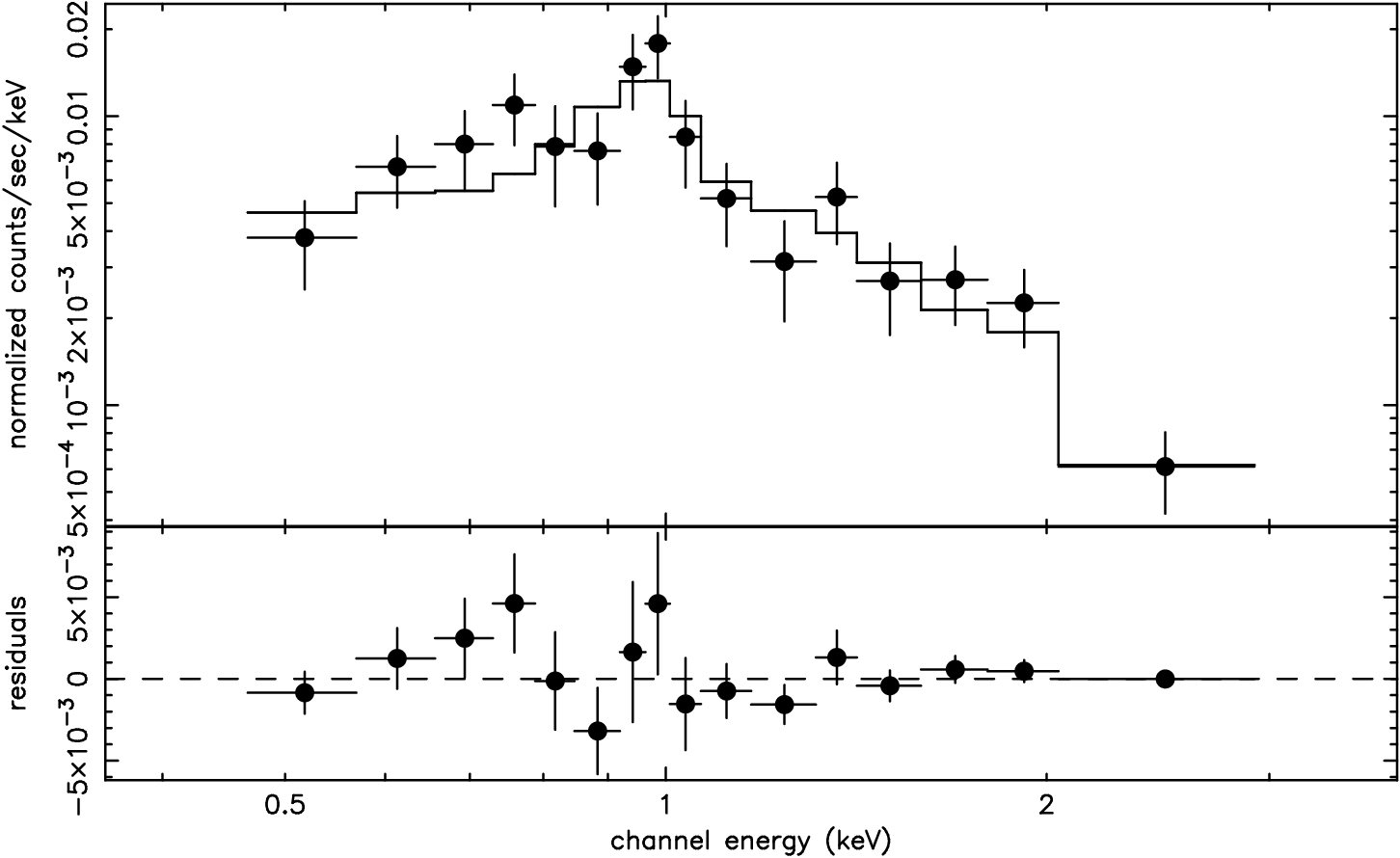}
\caption{Spectra of the 3 brightest discrete sources in NGC3256 in the 
0.5-10 keV range: source 7 (the Northern nucleus - top), 9 (centre)
and 10 (bottom). The top panels of each plot show the data and folded 
model (see Table 2). The bottom panels show the residuals to the fit.}
\end{figure}

A wavelet detection algorithm was used to search for compact sources
associated with NGC3256. This algorithm is particularly suited for the
detection of closely spaced and extended sources.  However, given the
strong and complex extended emission component in the central region
of NGC3256, it still is possible that strong fluctuations of the
background can be identified as spurious sources. For this reason, the
detection algorithm was run using an image in the 1.5-10 KeV energy
range only, where discrete sources are more conspicuous. Also, low
surface brightness and obviously extended sources in the Arm and to
the NW of the bright central source were disregarded since their
morphology did not appear less smeared in the hard band image.  The
final source list was compiled only after the determination of the
detection significance for each source. A total of 14 compact sources
were found with a detection significance (=(source counts)/(background
standard deviation) ranging from $\sim 140$ to $\sim 12$ (see Table
1). The sources are shown in Figure 1 and labeled from 1 to 14 with
increasing right ascension.

For each discrete source a background subtracted number of counts was
obtained from within a $r = 3.5$ pixel aperture (1.72 arcsec). This
encircles $\sim 99\%$ of the photons at 0.3 keV and $\sim 90\%$ of the
photons at 5.0 keV (in general, sources were not detected above this
energy).  The small apertures used imply that on average only $\sim2$
of the source counts are expected to correspond to the background.
Therefore, the study of these sources was done without screening out
the high background episodes. None of the sources was affected by
pileup. For each individual source a local background region was
defined by using an annulus with $r \sim 5 - 10$ pixels centered at
the position of the source with holes excised at the position of any
other nearby compact sources. The determined counts and associated
significance can be found in Table 1.

\subsection{Spatial Analysis}

We have analysed the spatial profile of the detected sources. To do
this we obtained locally background subtracted counts within circular
apertures of radius 1, 2, 3, 4 and 5 pixels and compared them with
model PSFs generated at the position of each source (Figure 2). The
full 0.3-10 keV band image was used to obtain the source profiles,
except for both nuclei, for which a harder 2-5 keV band image was
used. In this way the amount of diffuse emission around the northern
nucleus was minimised, and the detectability of the obscured southern
nucleus was maximised. The major difficulties in obtaining a
meaningful analysis of the profiles are the poor sampling of the PSF
(at 1.5 keV, for example, $\sim 90\%$ of the energy is expected to
fall within only 4 pixels), and the presence of the strong and
spatially variable diffuse component.

For isolated sources, such as sources 1, 4, 5, 12, 13 and 14, the
photometry suffers little from contamination by extended emission and
therefore the profiles are reliable despite the small number of counts
detected for some of the sources. All sources show profiles consistent
with the model PSF, although source 5 shows some evidence for an
extended component. However, as this is based on only one data point,
we will assume that all isolated sources are indeed unresolved.

The scenario is less straightforward for those sources located in the
innermost region of the galaxy since the background estimates will not
account for any spatial changes of the diffuse emission within the
apertures used to obtain the profiles. Figure 2 shows that while the
spatial distribution of sources 3, 6, and 8 (the southern nucleus) 11
and 13 are well described by the model PSF, sources 2, and 7 (the
northern nucleus) 9 and 10 are clearly resolved.

A simple Gaussian fit to the profiles shown in Figure 2 gives FWHM
values (the only parameter allowed to vary during the fitting) of
$\sim 0.8-1.5$ pixels ($\sim 0.4-0.7$ arcsec) for unresolved sources,
while extended sources had FWHM $\sim 2.2-2.8$ pixels (implying
deconvolved FWHM $\sim 1.8-2.5$ pixels or $\sim 0.9-1.3$ arcsec). The
model PSF had a FWHM $\sim 1.2$ pixels ($\sim 0.6$ arcsec). The only
source that did not follow this trend was source 2, which has a best
fit value of 1.5 pixels (ie, within the range of the unresolved
sources), even though Figure 2 suggests that the profile is extended.

The aperture used to perform the count subtraction ($r = 3.5$ pixels)
is only adequate if it contains most of the flux from the extended
sources. The Gaussian fit to the profiles shows that these sources
have a FWHM of up to 2.8 pixels (ie, $3\sigma = 3.6$ pixels or $\sim
1.8$ arcsecs). A well centred extraction aperture of 3.5 pixels,
therefore, will contain most of the source counts.  Since all extended
sources show a well defined peak, and sources 7, 9 and 10 correspond
to the brightest discrete sources found in the galaxy, the centering
of their flux distribution is well determined and, therefore, these
apertures were adequate.

\subsection{Fluxes and luminosities}

\begin{figure*} 
\centering
\includegraphics[angle=270,scale=0.7]{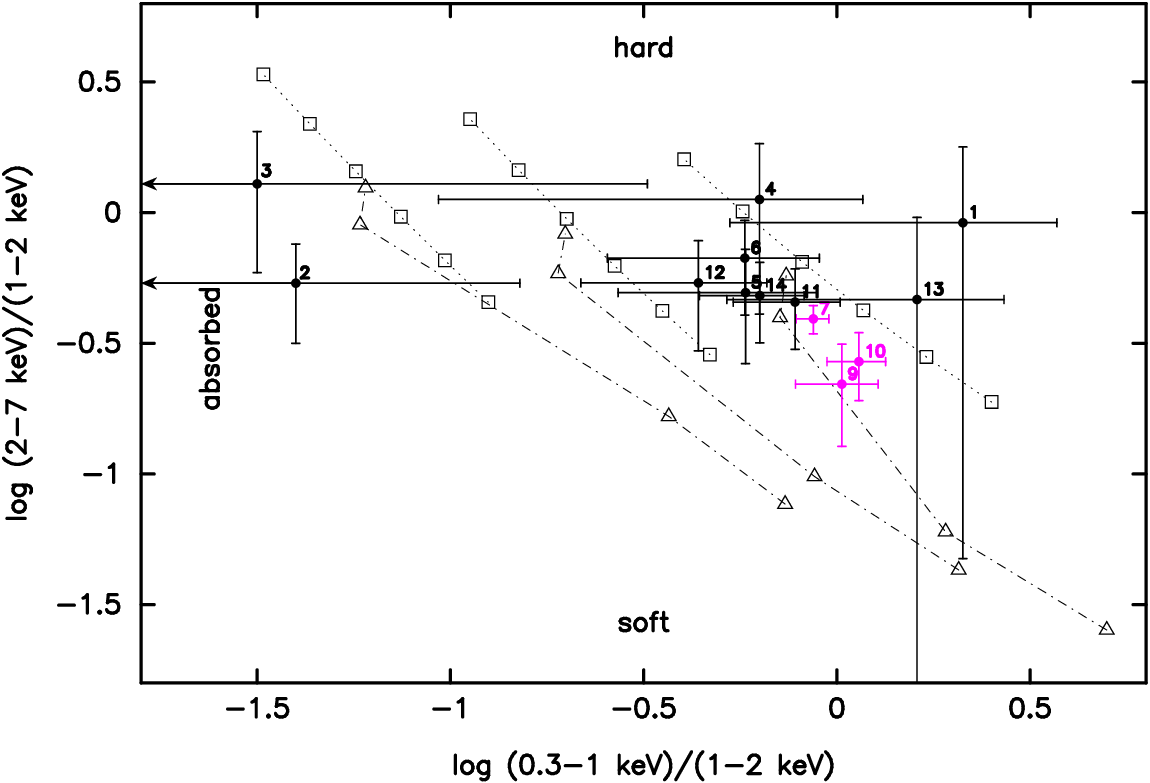}
\caption{`Hardness-ratios' (colour-colour) diagram for X-ray discrete sources.  
The X-ray colours were defined as the ratio of counts observed in the
0.3-1.0 keV, 1.0-2.0 keV and 2.0-7.0 keV energy bands. One $\sigma$
error bars are shown. The grids of models correspond to single
Raymond-Smith plasmas (triangles - dash-dotted line) with temperatures
(from top to bottom) 6.0, 3.0, 1.0 and 0.5 keV, and single power laws
(squares - dotted line) with index $\Gamma$ (from top to bottom) 0.5,
1.0, 1.5, 2.0, 2.5 and 3.0. From right to left successive grids
correspond to absorbing columns of $10^{21}$ (just above Galactic),
$5\times10^{21}$ and $10^{22}$ cm$^{-2}$ respectively. Sources 7, 9
and 10, for which spectral fits were properly derived are also
included.}
\end{figure*}

Only three of the 14 detected discrete sources had sufficiently high
numbers of counts to perform a useful spectral analysis. These
correspond to source 7, the northern nucleus, and sources 9 and
10. The spectra were modeled with a single (Mekal (M) or power-law
(PL)) component and a double (Mekal and power-law (M+PL) model in the
0.5-10 keV band pass. These models are representative of sources
dominated by thermal emission (such as SNRs and galactic bubbles),
non-thermal emission (such as accreting compact objects), and
composite objects.

For all sources the single Mekal model was a very poor representation
of the data, while the single power-law gave a much better fit,
particularly for source 9, given the low number of bins available (see
Table 2 and Figure 3). For sources 7 and 10, M+PL models were also
tried. For source 7 two independent hydrogen columns were adopted for
each spectral component in model $c$. This model represents a
physically plausible scenario with the low temperature Mekal component
representing a large-scale extended gaseous phase, and the harder,
more absorbed component arising from embedded or intrinsically
absorbed sources directly associated with the northern nucleus. Since
the best fit values of the hydrogen column affecting the low
temperature Mekal component were very close to or slightly less than
Galactic, we fixed the $N_{H}$ to the Galactic column (see note-foot
in Table 2), resulting in the same number of degrees of freedom as
model $b$.

To convert the count rates to fluxes for the remaining sources we must
assume an intrinsic spectral shape and intervening absorbing
column. In order to gain some insight into the best values to adopt,
we studied the X-ray colours of the sources, defined as the ratio of
counts observed in the 0.3-1.0 keV, 1.0-2.0 keV and 2.0-7.0 keV energy
bands. The bands were chosen in order to maximise the detection of the
sources in all three bands, as well as to obtain a good
characterization of the spectra. To compare the observed X-ray colours
of the sources with different spectral shapes, we computed the colours
of pure power-law and Raymond Smith models using the HEASRAC Portable,
Interactive, Multi-Mission Simulator (PIMMS - Mukai 1993).

Grids of points were determined for parameters $\Gamma = 0.5 - 3.0$
and $kT = 6.0 - 0.5$ keV, for the power-law and Raymond Smith models
respectively, and for $N_{H} = 10^{21} - 10^{23}$ cm$^{-2}$. In figure
4 we have plotted the computed grids along with the observed colours
of the discrete sources. Error bars were computed as 1 standard
deviation in the count ratios. The southern nucleus (source 8) is not
shown since it is not detected in either of the two softest bands used
in the diagram (see further discussion below in this section). 

The positions for sources 7, 9 and 10 can be used to investigate the
limitations in using these type of diagrams, since we have determined
their spectral parameters from spectral fits. From the single
power-law fits showed in Table 2 we find that all sources have an
absorbing column higher than Galactic but lower than $5 \times
10^{21}$ cm$^{-2}$, in agreement with the location of the sources in
the colour-colour diagram. However, from the diagram the columns are
underestimated for sources 7 and 10. For source 7 for example, $N_{H}
\sim 4 \times 10^{21}$ cm$^{-2}$ is found from the spectral fit, while the
column deduced from the colour-colour diagram is about $10^{21}$
cm$^{-2}$.  The fitted power-law indices also roughly agree with the
location of sources 9 and 10, giving values of $\Gamma \sim 2.5 -
3.0$. A larger discrepancy is seen for source 7 where the power-law
index estimated from the diagram is harder ($\Gamma \sim 2.0 - 2.5$)
than its fitted value. These disagreements could be due to the
observed spectral complexity of the emission in source 7 which
cannot be recovered by the over-simplified single parameter models
used in the colour-colour diagram. For source 9, however, the only
case where the single power-law model gives an acceptable fit
($\chi^{2}_{\rm red} \sim 1.0$), the best fit values suggest a hard
source and a moderate hydrogen column ($\Gamma \sim 2.4$, $N_{H}
\sim 1.6 \times 10^{21}$ cm$^{-2}$), which is not accurately reflected
by the position of the source in the colour-colour diagram. These
comparisons clearly illustrate the limits to using colour-colour and
hardness-ratio diagrams.

The positions of the remaining sources in the colour-colour diagram
show that most of the sources have an absorbing column above the
Galactic value, ie between $10^{21}$ and $5\times10^{21}$
cm$^{-2}$. The bulk of the sources show a harder spectral shape than
those for which we have performed a spectral fitting and therefore a
power law with index between $\Gamma = 2.0$ and 2.5 seems a reasonable
spectral shape to adopt. Bearing in mind the limitations in obtaining
spectral parameters from the colour-colour diagram as discussed above,
the derived power-law indices are in good agreement with that expected
from a population of high-mass X-ray binaries ($\Gamma \sim 1-2$
White, Nagase \& Parmar, 1995) or young supernova remnants ($kT
\sim 1-4$ keV, Bregman \& Pildis 1992). Sources 2 and 3, however, do
not follow the trend shown by the rest of the sources and present
evidence for much larger absorption.  These sources are not detected
in the 0.3-1.0 keV band and we only have an upper limit for their
(0.3-1.0 keV)/(1.0-2.0 keV) colour. A column density of $10^{22}$
cm$^{-2}$ was chosen as a representative value for these two heavily
absorbed sources.

\begin{table}
\centering
\caption{Fluxes and luminosities in the 0.5-10 keV energy range for the 
compact sources assuming a single power law component with index
$\Gamma=2.0$ and 2.5. The adopted hydrogen columns are also given. 
{\it Observed\/} fluxes are in units of $10^{-14}$  ergs s$^{-1}$ 
cm$^{-2}$ and {\it intrinsic\/} luminosities are in units of 
$10^{40}$ ergs s$^{-1}$ (for an assumed distance of 56 Mpc).}
\begin{tabular}[t]{lrcccc} \hline
        &               & \multicolumn{2}{c}{$\Gamma=2.0$} & \multicolumn{2}{c}{$\Gamma=2.5$} \\
Source & N$_{\rm H}$    & $F_{\rm x}$ & $L_{\rm x}$ & $F_{\rm x}$ & $L_{\rm x}$\\ \hline
1       &$10^{21}$      &0.43	&0.19	&0.30	&0.14\\
2       &$10^{22}$    	&2.12	&1.38	&1.59	&1.43\\
3       &$10^{22}$      &1.62	&1.05	&1.22	&1.10\\
4       &$5\times10^{21}$&0.81	&0.45	&0.59	&0.42\\
5       &$10^{21}$      &1.13	&0.49	&0.78	&0.36\\
6       &$5\times10^{21}$&2.91	&1.63	&2.20	&1.58\\
7(N)    &$10^{21}$      &13.5	&5.80	&9.25	&4.29\\
8(S)    &$5\times10^{22}$&3.00	&3.04	&2.47	&4.10\\
9       &$10^{21}$      &3.25	&1.39	&2.22	&1.03\\
10      &$10^{21}$      &4.96	&2.13	&3.40	&1.58\\
11      &$10^{21}$      &2.27	&0.97	&1.56	&0.72\\
12      &$5\times10^{21}$&1.52	&0.85	&1.10	&0.79\\
13      &$10^{21}$      &0.51	&0.22	&0.35	&0.22\\
14      &$5\times10^{21}$&2.78	&1.56	&2.02	&1.45\\[3pt]
Total   &               &40.81  &21.15  &29.05  &19.21\\
\hline
\end{tabular}
\end{table}

\begin{figure}
\centering
\includegraphics[angle=270,scale=0.3]{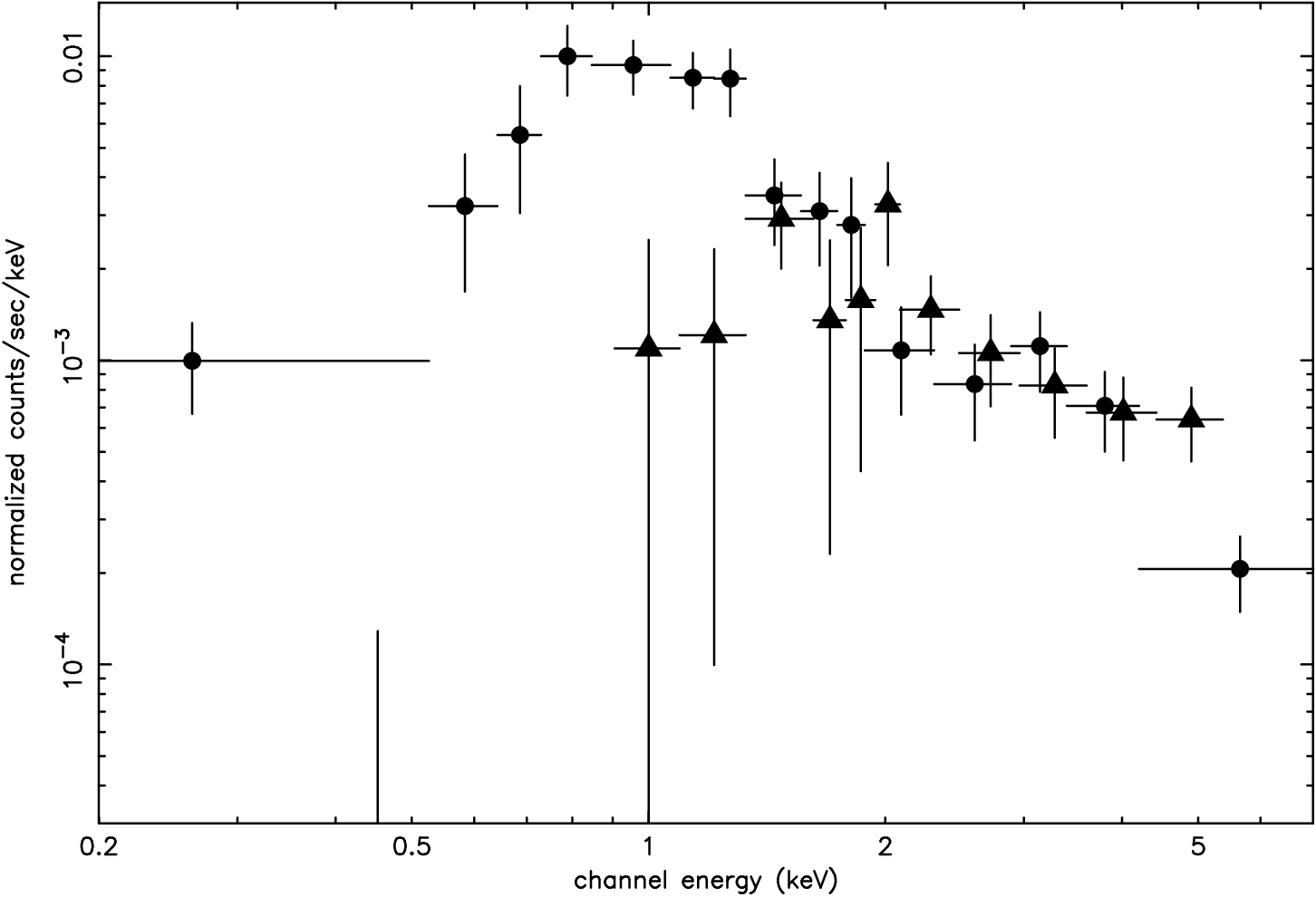} 
\caption{Stacked spectra for heavily absorbed sources 2, 3 and 8
(triangles), and less absorbed sources sources 5, 6, 11 and 13
(circles).}
\end{figure}

In order to check that the range of power-law indices and absorbing
columns inferred from the colour-colour diagram are representative of
`real' spectral shapes, we produced two `stacked' spectra for a group
of obscured and unobscured sources. The sources were selected based on
their observed colours in Figure 4 and from their location with
respect to the reddening observed in a colour image of the central
region of the galaxy (see section 6.2). Figure 5 clearly shows that
although the observed hard tails present a similar slope for the two
stacked spectra, the soft end shows clear evidence for significant
absorption in these sources, in good agreement with their position in
the colour-colour diagram. Single component models (M or PL) were
fitted to the stacked data and showed that the absorbing column
affecting the unobscured sources was a few times the Galactic value,
while the corresponding value for the obscured sources was about an
order of magnitude larger.

The southern nucleus is only detected at hard energies, and no counts
are detected below $\sim 2.0$ keV. This is perhaps not surprising,
since it is well established that this is a heavily obscured region.
To characterise the spectral shape of this nucleus we have again used
a colour-colour diagram, but this time using the 2.0-3.3 keV, 3.3-4.5
keV and 4.5-10 keV energy range. The bands were chosen in order to
maximise the background subtracted number of counts in each one of
them. We then compared the observed (2.0-3.3 keV)/(3.3-4.5 keV) and
(4.5-10 keV)/(3.3-4.5 keV) colours with grids of power law models as
described previously. We find that the spectral shape of the source is
well parameterized by a range of indices $\Gamma \sim 1.0 - 2.5$ if an
absorbing column of $5\times10^{22}$ cm$^{-2}$ is adopted. A higher
column ($10^{23}$ cm$^{-2}$) is possible but only if $\Gamma \ga 2.0$,
while a lower column ($10^{22}$ cm$^{-2}$) requires $\Gamma \la
1.5$. Columns above $10^{23}$ cm$^{-2}$ or below $10^{22}$ cm$^{-2}$
are not consistent with the observed colours.

\begin{figure}
\centering
\caption{Discrete source lightcurves. The average count rate for each source is
plotted as a straight line.}
\end{figure}

The final computed fluxes and luminosities for all sources are listed
in Table 3. They were computed assuming a power law spectrum with
indices $\Gamma = 2.0$ and 2.5 and the hydrogen columns discussed
above. The adoption of the different indices implies changes in the
fluxes of about $\sim 30\%$. Much larger errors could be introduced if
the spectral shapes are not well described by the assumed models, as
can be seen by comparing the fluxes and luminosities obtained for
source 7 from the spectral fitting and the adopted parameters used in
Table 3. Statistical errors will dominate, however, for the faintest
sources ($\la 30$ counts - see Table 1). Another significant source of
uncertainty is the adoption of a representative background. This
problem is particularly important for sources embedded in strong
diffuse emission. In all, the fluxes and luminosities in Table 3 are
probably good to within a factor of two or three.  Inspection of
Tables 1 and 3 shows that our sensitivity threshold is $\sim 4 \times
10^{-15}$ ergs s$^{-1}$ cm$^{-2}$ ($\sim 1.5 \times 10^{39}$ ergs
s$^{-1}$) for an isolated source, and $\sim 1.5 \times 10^{-14}$ ergs
s$^{-1}$ cm$^{-2}$ ($\sim 6 \times 10^{39}$ ergs s$^{-1}$) for sources
embedded within the diffuse emission.

\subsection{Time variability}

\begin{figure*}
\centering
\includegraphics[angle=270,scale=0.35]{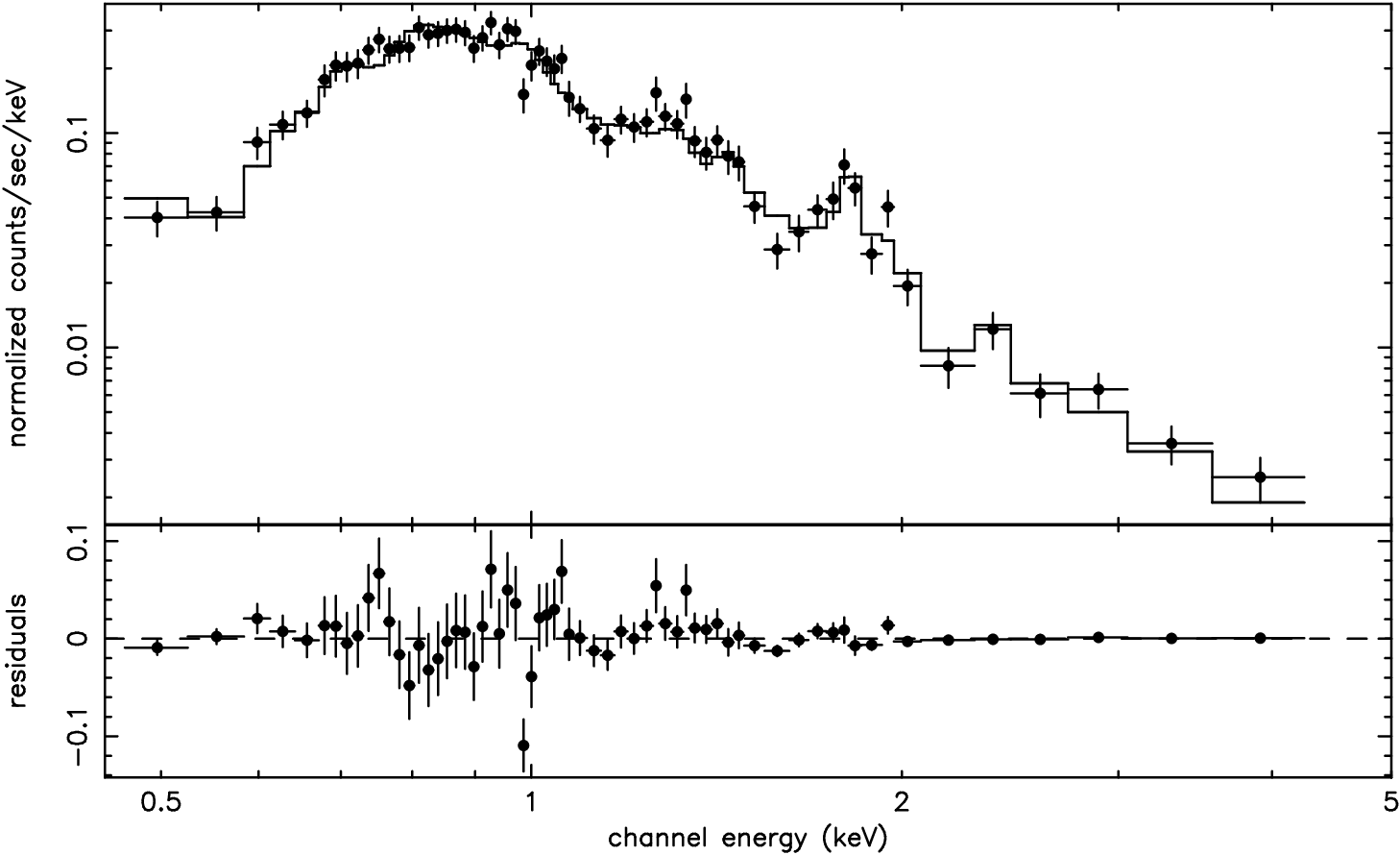}\hspace{0.5cm}
\includegraphics[angle=270,scale=0.33]{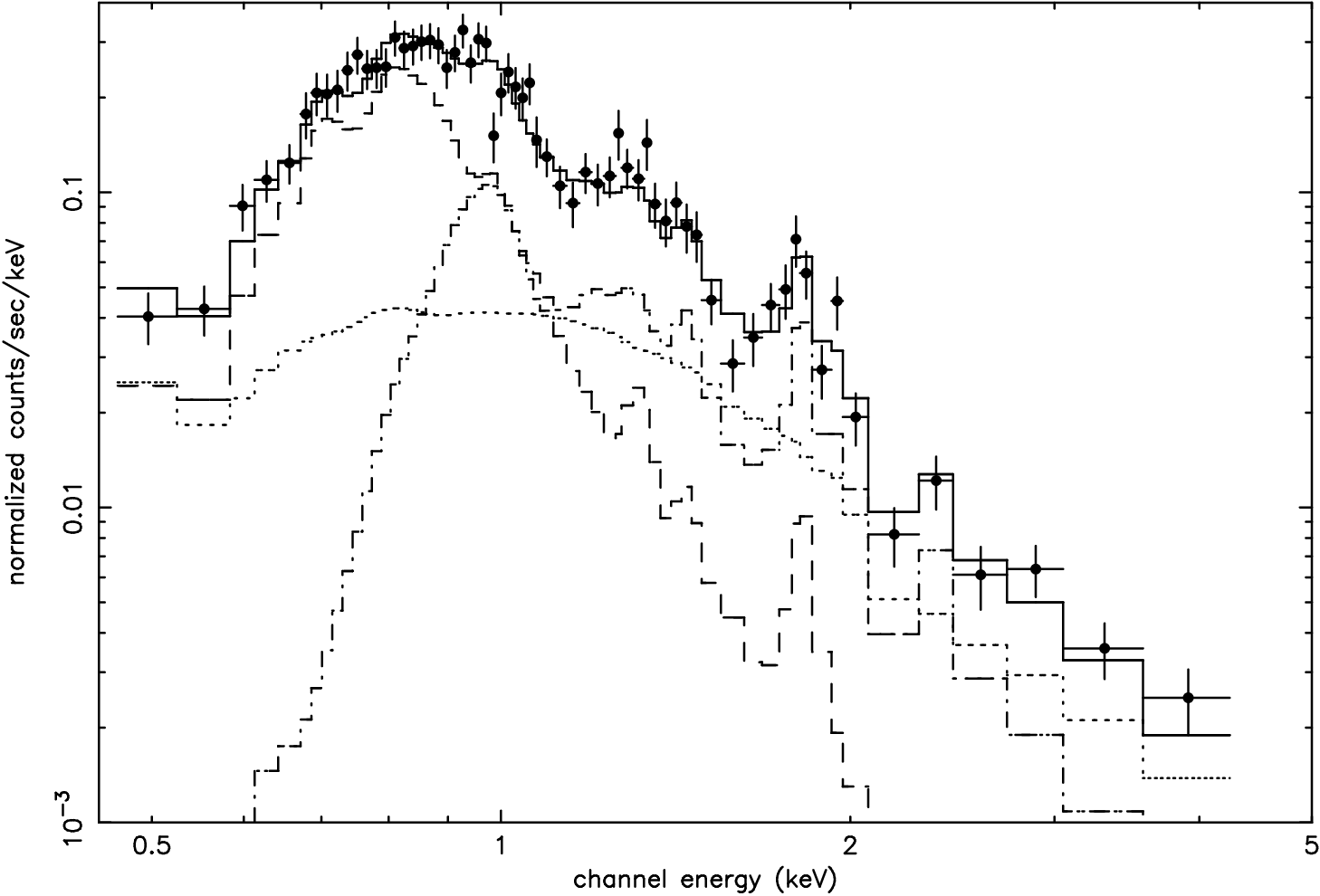}
\caption{Left: Spectrum of the diffuse component in NGC3256 in the 
0.5-5 keV range. The top panel shows the data and folded model; the bottom 
panel shows the residuals in the same energy range. Right: best fit model 
components.}
\end{figure*}

In order to search for time variability of the discrete sources light
curves for the 6 brightest objects were constructed using 6 bins of
4.3 ksec each, spanning a total of 26 ksecs. The errors associated
with the count rates correspond to 1$\sigma$ deviations assuming
Poisson statistics. The light curves are shown in Figure 6.

A superficial examination shows no obvious signs of variability in the
lightcurves. We carried out a $\chi^{2}$ test to compare the
lightcurves with a constant distribution. The test showed that all
distributions are consistent with no variability ($\chi^{2}_{\rm red}
\la 0.5$), except for source 6 where no variation is a poorer 
representation of the data ($\chi^{2}_{\rm red} = 1.59$). However,
this is not a significant result, and the null hypothesis is not
rejected at the 85\% confidence level.

\section{Diffuse emission}

\subsection{Integrated spectrum}

\begin{table*}
\centering
\caption{Best-fit model parameters for the diffuse emission in the
0.5-10 keV energy range. Solar abundances were assumed for the Mekal
components. Intervening hydrogen columns are in units of $10^{21}$
cm$^{-2}$ and plasma temperatures are in units of keV. $Gal$ stands
for a foreground Galactic hydrogen column ($9 \times 10^{20}$
cm$^{-2}$). Errors correspond to 90\% CL for one parameter of
interest. {\it Observed\/} fluxes are in units of $10^{-13}$ ergs
s$^{-1}$ cm$^{-2}$ and {\it intrinsic\/} luminosities are in units of
$10^{41}$ ergs s$^{-1}$ (for an assumed distance of 56 Mpc).}
\begin{tabular}{lccccccccr} \hline
Model & $N_{H}$ & $kT$                   & $N_{H}$             & $kT$                   & $N_{H}$             & $kT$ or $\Gamma$        & $\chi^{2}_{\rm red}$(dof) & $F_{x}$ & $L_{\rm x}$ \\ \hline
M+M+M   & $Gal$   & $0.37^{+0.04}_{-0.03}$ & $7.1^{+1.1}_{-1.0}$ & $0.76^{+0.04}_{-0.05}$ & $\dag$              & $2.76^{+1.90}_{-0.67}$  & 1.54(59) & 6.78 & 8.32\\[5pt]
M+M+PL  & $Gal$   & $0.54^{+0.18}_{-0.13}$ & $4.6^{+1.5}_{-1.2}$ & $0.89^{+0.13}_{-0.19}$ & $\dag$              & $3.03^{+0.25}_{-0.49}$  & 1.45(59) & 6.62 & 5.73\\[5pt]
M+M+M   & $Gal$   & $0.60^{+0.04}_{-0.05}$ & $9.7^{+1.8}_{-1.6}$ & $0.90^{+0.10}_{-0.07}$ & $Gal^{+0.8}\ddag$   & $3.93^{+4.54}_{-1.14}$  & 1.36(58) & 6.83 & 7.23\\[5pt]
M+M+PL$^{\star}$&$Gal$& $0.60^{+0.04}_{-0.07}$&$9.2^{+2.2}_{-2.2}$&$0.91^{+0.14}_{-0.09}$ & $3.4^{+1.3}_{-1.8}$ & $2.75^{+0.49}_{-0.68}$  & 1.31(58) & 6.60 & 6.79\\
\multicolumn{10}{l}{}\\
\multicolumn{10}{l}{$\dag$ Same hydrogen column as applied to previous spectral component.}\\[3pt]
\multicolumn{10}{l}{$\ddag$ The best fit value of $N_{H}$ was found to be slightly under Galactic.}\\[3pt]
\multicolumn{10}{l}{$^{\star}$ Model component normalizations: $1.1\times10^{-4}$ and $4.9\times10^{-4}$ for the 0.6 and 0.9 keV Mekal components}\\ 
\multicolumn{10}{l}{respectively, in units of $(10^{-14}/(4 \pi D^{2})) \times$ $\int n_{e}^{2}dV$ (where $D$ is the distance to NGC3256 in cm, $n_{e}$ is the}\\  
\multicolumn{10}{l}{electron density in cm$^{-3}$, and $V$ is the volume of the emitting region in cm$^{3}$); $1.5\times10^{-4}$ for the Power}\\
\multicolumn{10}{l}{Law, in units of $10^{-4}$ photons cm$^{-2}$ s$^{-1}$ keV$^{-1}$.}\\
\hline
\end{tabular}
\end{table*}

In order to perform a spectral analysis of the diffuse component in
NGC3256 a large aperture with a diameter of 40 arcsec ($\sim 11$ kpc)
was defined after removing all discrete sources. At this distance from
the central peak of emission the diffuse emission had a comparable
surface brightness to the background. The large aperture used implies
that the observed episodes of high background described in Section 2
could contaminate the intrinsic emission from the galaxy. Therefore
the data were filtered to exclude background count rates above 1.5
counts per second.  Consequently only 14.5 ksec ($\sim 52\%$) of the
total observing time was used. These screened data are used for all
the analysis described here and in section 4.2.

The aperture was fully contained within only one (out of four) of the
readout nodes of the S3 chip (node to node sensitivity changes can be
as large 10 percent). In fact, the bulk of the extended component
covered a roughly circular region with a diameter $\la 30$ arcseconds,
compact enough for the response of the detector to remain fairly
constant. For this reason only one set of Fits Embedded Functions (FEF
- a FITS table which stores the mapping of the energy to pulse height)
was used to calibrate the spectrum. (Using a response built up from
the weighted combination of the 3 to 4 FEFs that fully covered the
area subtended by the diffuse component did not change the fit to the
data). The background was obtained from a large semi-annular region
surrounding NGC3256 ($r \sim 75 - 115$ arcsec) and fully contained
within the same readout node as the emission from the galaxy.

Figure 7 shows the spectrum of the diffuse emission. A total of $\sim
2860$ counts were detected within the aperture, of which at most $\sim
100$ could be from the background. The relative softness of the
emission is evident as no counts are detected above $\sim 5$ keV. The
presence of strong emission lines is also readily visible at $\sim
1.4$ keV and $\sim 1.8$ keV. At these energies we expect contributions
from lines such as He-like and K-shell emission of Mg and Si ions,
indicating the presence of a thermal component with $kT < 2$ keV.

The spectrum of the diffuse emission was first fitted using single and
double component models which yielded unacceptable chi-squared
values. We then followed with three-component models (M+M+M or
M+M+PL). The presence of multi-temperature gas has already been
claimed for several starburst galaxies, such as NGC1808 (Awaki etal
1996), Arp 299 (Heckman etal 1999), NGC4449 (della Ceca, Griffiths \&
Heckman, 1997), NGC253 and M82 (Ptak etal 1997). The introduction of a
power law component could account for an unresolved population of
accreting binaries. Since different spectral components can have
different degrees of absorption (see references above), we allowed for
different $N_{H}$ values for the fitted components. We also fixed the
hydrogen column absorbing the softest thermal plasma component to the
Galactic value, since fitting the parameter nearly always resulted in
a Galactic or slightly under-Galactic value.

The results of the spectral fits are reported in Table 4. Solar
metallicities were assumed throughout, consistent with the optically
measured abundances (L\'{\i}pari etal 2000). If left free to vary, the
abundances imply a solar or over-solar value, without significantly
improving the fit. For the first two models in Table 4, the second and
third spectral components are subject to the same hydrogen column. For
the third and fourth model, each component has an independent
intervening $N_{H}$.  Despite the clear improvement in the statistics
of the successive models presented in Table 4, it was impossible to
reduce $\chi^{2}_{\rm red}$ below 1.3. This is probably due to the
still preliminary status of the instrument responses available for the
calibration of the data. The uncertainties mostly affect the modeling
of line emission dominated spectra. All the models in Table 4 are
consistent with a general picture of a moderately absorbed cool
thermal plasma plus a warmer thermal component with a much higher
absorption.  A steep power law or a thermal component with a
temperature of about 4 keV is necessary to explain the faint hard tail
seen in the spectrum. Table 5 lists fluxes and luminosities derived
for each component from the model with the best reduced chi-squared
(M+M+PL model; last entry in Table 4).

\begin{table}
\centering
\caption{Fluxes and luminosities for the diffuse emission in the 0.5-10 
keV energy range from the M+M+PL model. {\it Observed\/} fluxes are in 
units of $10^{-14}$ ergs s$^{-1}$ cm$^{-2}$ and {\it intrinsic\/} 
luminosities are in units of $10^{41}$ ergs s$^{-1}$ (for an assumed 
distance of 56 Mpc).} 
\begin{tabular}{lcccc} \hline
Component 	& $F_{\rm x}$ & $L_{\rm x}$& $F_{\rm x}$ & $L_{\rm x}$ \\ 
  	 	& 0.5-2.5     & 0.5-2.5    & 2.5-10      & 2.5-10  \\
  	 	& (keV)       & (keV)      & (keV)       & (keV) \\ \hline
Cool Mekal	& 20.5        & 1.31       & 0.21        & 0.01 \\
Warm Mekal	& 17.2        & 3.83       & 3.07        & 0.13 \\
Power Law	& 15.1        & 1.38       & 9.96        & 0.38 \\
Total		& 52.8        & 6.26       & 13.2        & 0.52 \\
\hline
\end{tabular}
\end{table}

\subsection{Spatially resolved spectra}

The spectral-imaging capabilities of {\it Chandra\/} enable a more
detailed study of the diffuse emission. Ideally, we would like to
identify each of the spectral components determined in the previous
section with a specific spatial region, but this is only possible if
each component strongly dominates the local emission.

Two thermal components have been clearly identified in the {\it
Chandra\/} data. Observational evidence on galactic winds suggests
that the central regions of starbursts are characterised by higher
plasma temperatures than the outer regions (Strickland, Ponman \&
Stevens 1997, Pietsch etal 2000). This is also supported by various
theoretical models of galactic winds (Suchkov etal 1996, Chevalier \&
Clegg 1985). Given the nearly face-on orientation of NGC3256 we
decided to extract spectra from annular regions centered in between
both nuclei. After experimenting with different radii we defined three
regions, A, B and C with outer radii of 15, 23 and 40 pixels ($\sim 4,
6,$ and 11 kpc). This choice gives a similar number of counts ($\sim
900-1000$) in each region and also provides a good representation of
the spectral changes with radius. Region C is roughly coincident
(although with a larger outer radius) with the onset of a ring of
broad H$\alpha$ line emission (FWZI up to $\la 6000$ km s$^{-1}$)
probably associated with the wind, as discussed by MLH.

The regions are shown in Figure 8 overlaid onto a smoothed image of
the X-ray emission. The spectra obtained for each region (after
removing the discrete sources) are also shown. The spectra show that
while the emission above 1 keV shows a similar trend for all radii,
the dispersion is much larger below 1 keV, suggesting that different
amounts absorption and/or peak temperatures may characterise the
different regions.

In order to quantify the different spectral trends observed in Figure
8 we fitted each spectrum with a thermal plasma plus a power law
(M+PL), each component absorbed by an independent hydrogen column. The
power law component was fixed with an index of 2.7. The results are
shown in Table 6. The best fitted value for the hydrogen column
affecting the thermal plasma in region C was very close to Galactic
and therefore, given that a column much larger than Galactic is
unlikely to pertain across such an extended region, a value of $Gal$
was adopted.

\begin{figure*}
\centering
\begin{minipage}[c]{0.5\textwidth}
\centering \includegraphics[scale=0.46]{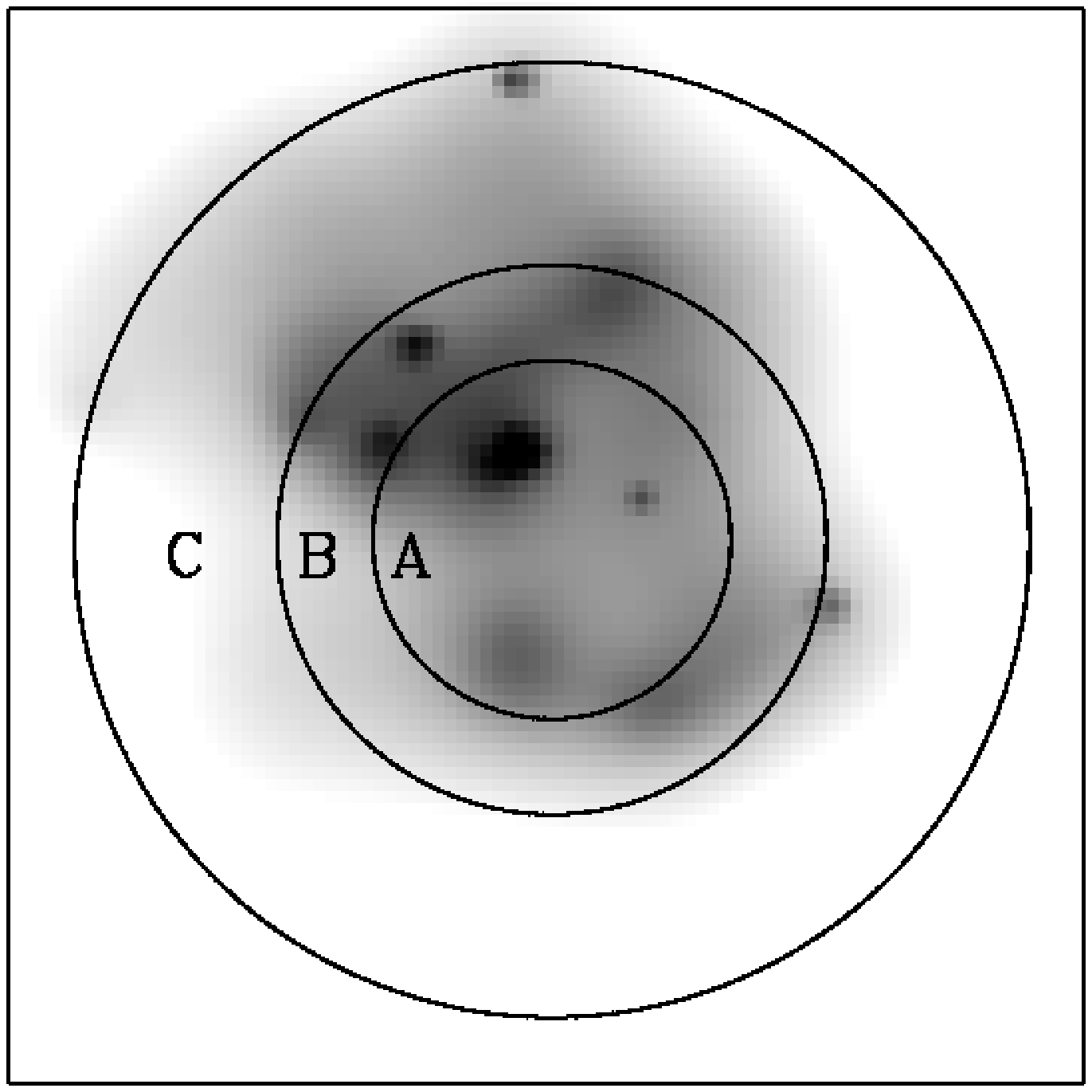}%
\end{minipage}%
\begin{minipage}[c]{0.5\textwidth}
\centering \includegraphics[angle=270,scale=0.4]{fig8b.ps}
\end{minipage}
\caption{Left: Smoothed X-ray image in the 0.3-1.5 keV energy range
showing the regions defined for the spatially resolved study of the
diffuse emission. The radii of the regions correspond to $\sim 4, 6,$
and 11 kpc. Right: Spectra of the three regions in the 0.5-10 keV
energy range.}  
\end{figure*}

\begin{table}
\centering
\caption{Best fit parameters for the diffuse emission from regions A, B and C.
A M+PL model was used with each component being absorbed by an independent $N_{H}$. 
The power law component was frozen to a value of 2.7. Hydrogen columns are 
given in units of $10^{21}$ cm$^{-2}$ and plasma temperatures in units of keV. 
The errors correspond to 90\% CL for 1 parameter of interest. {\it Observed\/} 
(ie, uncorrected) fluxes in the 0.5-10 keV for the thermal component only are 
given in units of $10^{-14}$ ergs s$^{-1}$ cm$^{-2}$.}
\begin{tabular}{lrrrrrr} \hline
Reg  & $N_{H}$             & $kT$                   & $N_{H}$             & $\Gamma$ & $\chi^{2}_{\rm red}$(dof) & $F_{\rm x}$ \\ \hline
A	& $7.7^{+2.2}_{-2.4}$ & $0.76^{+0.10}_{-0.16}$ & $2.3^{+0.7}_{-0.6}$ & 2.7	&1.08(29)  &  8.95 \\[3pt]
B	& $2.8^{+1.5}_{-1.4}$ & $0.60^{+0.05}_{-0.09}$ & $4.3^{+1.6}_{-0.9}$ & 2.7	&1.14(30)  &  8.61 \\[3pt]
C	& $Gal$               & $0.63^{+0.04}_{-0.04}$ & $3.6^{+2.1}_{-0.9}$ & 2.7	&1.18(34)  &  10.25 \\
\hline
\end{tabular}
\end{table}

The results in Table 6 show a trend of the spectral parameters with
radius. The most robust result concerns the amount of extinction
affecting each region, with $N_{H}$ varying by about an order of
magnitude between the central and the outer parts of the diffuse
emission.  The variation in temperature is less certain given the
associated errors. However, the presence of a $\sim 0.9$ keV highly
absorbed plasma has been well established by {\it ASCA} observations
(see section 5). So the correlation of higher temperatures with higher
$N_{H}$ supports this result.  Further more, we also extracted a
spectrum from a smaller ($r = 10$ pixels, $\sim 3$ kpc) circular
region which is not shown in Figure 8. The fit to the data (using the
same model as for the other regions) gives $N_{H}=11^{+6}_{-3} \times
10^{21}$ cm$^{-2}$ and $kT = 0.9^{+0.2}_{-0.3}$ keV for the column and
temperature of the thermal plasma, respectively. The large errors
reflect the poor statistics (only 13 bins were obtained;
$\chi^{2}_{\rm red}=0.9$) but it again supports the trend found in
Table 6.

\begin{table*}
\centering
\caption{Best-fit model parameters for the total emission in the
0.5-10 keV energy range. Solar abundances were assumed for the Mekal
components. Intervening hydrogen columns are in units of $10^{21}$
cm$^{-2}$ and plasma temperatures are in units of keV. $Gal$ stands
for a foreground Galactic hydrogen column ($9 \times 10^{20}$
cm$^{-2}$). Errors correspond to 90\% CL for one parameter of
interest. {\it Observed\/} fluxes are in units of $10^{-13}$ ergs
s$^{-1}$ cm$^{-2}$ and {\it intrinsic\/} luminosities are in units of
$10^{41}$ ergs s$^{-1}$ (for an assumed distance of 56 Mpc).}
\begin{tabular}{lccccccccr} \hline
Model & $N_{H}$ & $kT$                   & $N_{H}$             & $kT$                   & $N_{H}$             & $kT$ or $\Gamma$        & $\chi^{2}_{\rm red}$(dof) & $F_{x}$ & $L_{\rm x}$\\ \hline
M+M+PL            & $Gal$   & $0.50^{+0.12}_{-0.17}$ & $3.7^{+0.5}_{-0.7}$ & $0.88^{+0.21}_{-0.14}$ & $\dag$              & $2.42^{+0.17}_{-0.21}$  & 1.16(80) & 11.72 & 7.67\\[5pt]
M+M+PL$^{\star}$  & $Gal$   & $0.58^{+0.05}_{-0.14}$ & $7.8^{+1.7}_{-2.1}$ & $0.92^{+0.12}_{-0.17}$ & $3.4^{+1.3}_{-1.8}$ & $1.99^{+0.32}_{-0.36}$  & 1.07(79) & 12.09 & 9.09\\
\multicolumn{10}{l}{}\\
\multicolumn{10}{l}{$\dag$: Same hydrogen column as applied to previous spectral component.}\\[3pt]
\multicolumn{10}{l}{$^{\star}$ Model component normalizations: $1.3\times10^{-4}$ and $5.8\times10^{-4}$ for the 0.6 and 0.9 keV Mekal components}\\ 
\multicolumn{10}{l}{respectively, in units of $(10^{-14}/(4 \pi D)) \times \int n_{e}^{2} \ dV$ (where $D$ is the distance to NGC3256 in cm, $n_{e}$ is the}\\  
\multicolumn{10}{l}{electron density in cm$^{-3}$, and $V$ is the volume of the emitting region in cm$^{3}$); $1.8\times10^{-4}$ for the Power}\\
\multicolumn{10}{l}{Law, in units of $10^{-4}$ photons cm$^{-2}$ s$^{-1}$ keV$^{-1}$.}\\\hline
\end{tabular}
\end{table*}

The quantitative results presented in Table 6 should be treated with
caution. For example, the derived 0.5-10 keV flux for the central
region (A) accounts for less than half of the flux determined for the
0.9 keV component (see Table 5), which is thought to dominate in this
region. The use of over-simplified and poorly constrained models to
characterise the different emitting regions is probably responsible
for part of the disagreement. However, it is also clear that the two
thermal components observed in the integrated spectrum of the diffuse
emission are not representative of the spectral complexity present on
small scales; they only characterise the average properties of the
emission. It is not possible to explore these important issues further
with the present data, but a significantly longer exposure would no
doubt answer some of the open questions.

\section{Comparison with {\it ASCA\/} observations}

MLH have reported {\it ASCA\/} observations of NGC3256 and determined
a total luminosity for the system of $\sim 1.6 \times 10^{42}$ ergs
s$^{-1}$ in the 0.5-10 keV energy range. In comparison, Tables 2 and 3
imply a total luminosity of $\sim 9 \times 10^{41}$ ergs s$^{-1}$ in
the same spectral range as determined with {\it Chandra\/}, ie, a
reduction of $\sim 40\%$ in luminosity. We should attempt to
understand the origin of this difference.

Given the extremely low spatial resolution of the {\it ASCA\/} (PSF
FWHMS $\sim 3$ arcmin), the spectral analysis presented by MLH
corresponds to the integrated emission (ie, diffuse plus compact
sources) of most of the galaxy. Their adopted model consists of two
thermal plasmas ($kT \sim 0.3, 0.8$ keV) and a power law with photon
index $\sim 1.7$. The power law and the high temperature thermal
component are affected by a single absorbing column of $\sim 8 \times
10^{21}$ cm$^{-2}$. This model, however, is not a good description of
a {\it Chandra\/} spectrum obtained from the total emission observed
within the central 40 arcseconds of the galaxy. A model using the best
fit values derived by MLH gives an unacceptable fit ($\chi^{2}_{\rm
red} = 1.7$). In fact, if the parameters are allowed to vary (and a
Mekal component is used to described the thermal emission instead of a
Raymond-Smith plasma, as used by MLH), the best fit values for the
{\it Chandra\/} `total spectrum' are very similar to those found for
the diffuse emission only (compare Table 4 and Table 7), although a
flatter power law is determined for the total spectrum, as expected
from the hard spectrum that characterizes the discrete sources. From
this fit we derive an {\it observed\/} flux of $\sim 1.2 \times
10^{-12}$ ergs s$^{-1}$ cm$^{-2}$ in the 0.5-10 keV energy band, which
differs only by $\sim 5\%$ from the flux derived from the {\it ASCA\/}
observations.

The difference in the models adopted in the analysis of the {\it
ASCA\/} and {\it Chandra\/} data can explain the different derived
{\it intrinsic\/} fluxes and luminosities. The change is primarily due
to the reduction in the intrinsic emission (or component
normalization) of the warm, 0.8-0.9 keV component, as seen by {\it
Chandra\/}, which is the most luminous component in the model derived
by MLH. Inspection of Figure 7b in MLH, clearly shows a warm component
which is dominant in the 1-2 keV range, since the contribution from a
heavily absorbed power law component is almost negligible at these
energies. The {\it Chandra\/} data, on the other hand, is consistent
with a less absorbed and stronger power law component, implying a
smaller contribution from the warm thermal component to the overall
emission.

\section{Discussion}

\subsection{The double nuclei}

\subsubsection{Optical, IR and X-ray properties}

Optical and near-IR observations show that the properties of the
northern nucleus of NGC3256 correspond to those of a strong starburst.
L\'{\i}pari etal (2000) found that the optical emission line features
correspond to `those of low-ionization and high metallicity giant HII
regions'. On the other hand, the outer regions ($R > 10$ arcsec) are
characterised by shock excitation (L\'{\i}pari etal 2000;
MLH). L\'{\i}pari etal also presented UV spectra of the central region
($10\times20$ arcsec) showing strong absorption lines, implying the
presence of massive young stars. Kotilainen etal (1996) obtained
near-IR emission line maps of the centre of the galaxy and determined
an extinction towards the northern nucleus of $A_{V} \sim 2.4$
magnitudes.

The southern nucleus of NGC3256 was first imaged in the near-IR
continuum by Moorwood \& Oliva (1994), although near-IR spectroscopy
of the source had already been presented by Doyon, Joseph \& Wright
(1994). From their K band imaging Moorwood \& Oliva detected an
obscured source located 5 arcsec to the south of the central peak (the
northern nucleus) and suggested that this secondary peak corresponded
to the nucleus of the merging companion galaxy. Support came with
radio observations reported by Norris \& Forbes (1995) showing two
equally bright knots of emission. Near-IR observations by Kotilainen
etal (1996) showed that the luminosity of both nuclei is comparable in
the L (3.5 $\mu$m) band and they estimated an extinction towards the
southern nucleus of $A_{V} \sim 10$ magnitudes. High resolution N-band
(12 $\mu$m) images show, however, that the northern nucleus is $\sim
20$ times brighter than the southern peak in the mid-IR, suggesting
that the northern nucleus is the dominant star-forming region in the
galaxy (Lira \& Ward 2002).

\begin{figure*} 
\centering
\includegraphics[scale=0.8]{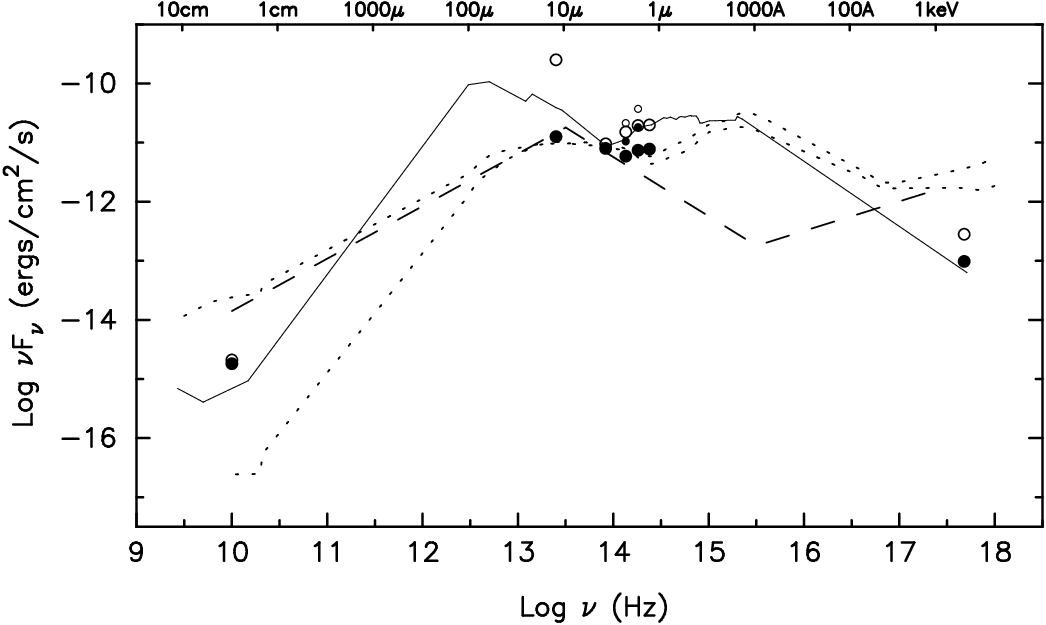} 
\caption{Absorption corrected fluxes for the northern nucleus (open 
circles) and southern nucleus (filled circles). Radio and IR data from
Norris \& Forbes (1995), Lira \& Ward (2002), and Kotilainen etal
(1996); X-ray data from this paper. {\it NICMOS\/} HK observations
(this paper) are shown with smaller size symbols. The data are
compared with the spectral energy distributions of radio loud and
radio quiet classical quasars (doted lines), low-luminosity AGN
(dashed line) and starburst galaxies (solid line). All distributions
have been normalized to the nuclear fluxes in the L (3.5 $\mu$m)
band.}
\end{figure*}

In order to improve the reddening estimations towards the galaxy
nuclei, we have analysed archive {\it NICMOS\/} data of the central
region of NGC3256. These observations were obtained on the 28$^{th}$
of November 1997 using the NIC2 camera (FOV $= 19.2 \times 19.2$
arcsec, FWHM for a point source $\sim 0.15$ arcsec) and the F222M
($\sim$ K band), F160W ($\sim$ H band), and F190N (Pa$\alpha$ plus
continuum at 1.90 $\mu$m) filters.  For details of the reduction
procedure see Alonso-Herrero etal (2000). Gaussian fitting to the
K-band data shows that both nuclei are resolved, with the northern
nucleus (deconvolved FWHM $\sim 0.25$ arcsec) being more compact than
the southern nucleus (deconvolved FWHM $\sim 0.34$ arcsec). Using an
aperture of 0.76 arcsec in diameter we measured an $H-K$ colour of
0.64 magnitudes for the northern nucleus and 1.44 magnitudes for the
southern nucleus. Assuming a intrinsic colour $H-K \sim 0.22$ for the
nuclei of starburst and Seyfert galaxies (Glass \& Morwood 1985) and
the extinction law of Cardelli, Clayton \& Mathis (1989) this implies
a visual extinction of $\sim 5.5$ and $\sim 16$ magnitudes for the
northern and southern nuclei respectively.  Assuming a Galactic
gas-to-dust ratio, these values of $A_{V}$ correspond to a hydrogen
column density for the northern nucleus of $\sim 8\times10^{21}$
cm$^{-2}$ and $\sim 2.5 \times10^{22}$ cm$^{-2}$ for the southern
nucleus, in good agreement with the values inferred from the X-ray
observations.

The northern nucleus is the brightest X-ray source in NGC3256 with an
unabsorbed luminosity of $\sim 8-10 \times 10^{40}$ ergs s$^{-1}$ in
the 0.5-10 keV energy range (see Table 2). The source is clearly
resolved, and therefore it probably corresponds to a compact grouping
of individual sources and diffuse emission at the heart of the nuclear
starburst. The deconvolved FWHM of the radial profile in the 2-5 keV
range shown in Figure 2 corresponds to $\sim 1$ arcsec, or $\sim 270$
pc at the distance of NGC3256. The full band, 0.5-10 keV profile has a
deconvolved FWHM of $\sim 1.5$ arcsec, corresponding to 400 pc, which
is only a factor $\sim 1.5$ times smaller than the central
starbursting region present in M82. The X-ray luminosity of both
regions are also similar. In the case of M82 the 2-10 keV emission is
dominated by a highly variable and extremely luminous point source
(Kaaret etal 2001, Ward etal 2001).

The NICMOS data for NGC3256 do not show conspicuous sources within a
radius of 1.5 arcsec of the northern nucleus. The FWHM of the peak is
only $\sim 0.25$ arcsecs or $\sim 70$ pc. Althought there is little
evidence for a one-to-one correlation between X-ray and IR point
sources, we would expect the overall size of these regions to be
similar (eg, M82 - Griffiths etal 2000). Therefore any population of
X-ray emitting discrete sources (accreting binaries and compact SNRs)
should be mainly associated with the very compact peak seen in the
NICMOS data. It seems unlikely that a population of `normal' binaries
and SNRs could explain the observed X-ray emission, since several
hundreds such sources would need to be packed within a very small
volume. A more plausible scenario is that the X-ray emission is
dominated by a diffuse component (eg, the base of the galactic wind -
support for this scenario comes from the rather steep index of the
power-law component determined in Table 2 which can also be fitted by
a hot gas component with $kT \sim 2$ keV), and/or by a few nuclear
ultraluminous sources.

\subsubsection{A hidden AGN?}

Many authors have searched for evidence of a hidden AGN in the heavily
obscured southern nucleus of NGC3256. An active nucleus could play an
important role in the energetics of the galaxy. So far observations at
IR and radio wavelengths have failed to provide unambiguous evidence
for the presence of an AGN (Kotilainen etal, 1996; Norris \& Forbes,
1995).  Mid-IR {\it ISO\/} spectroscopy of NGC3256 (Rigopoulou etal,
2000) also failed to detect the high-excitation emission lines
indicative of the presence of an AGN. The large aperture of {\it
ISO\/} may have completely diluted the emission from a hidden active
nucleus, but even if this is the case it would imply an intrinsically
weak AGN which would not make a significant contribution to the total
emission from the galaxy.

Our detection of the southern nucleus in X-rays allows us to address
the AGN hypothesis from a different perspective. The unresolved nature
of the source lends some support to the AGN scenario, although the
small number of counts, coupled with the strong absorption, could
conspire to hide any signature of an extended component.  Adopting a
power law spectral model with $\Gamma = 2.0$ and an absorbing column
of $5\times10^{22}$ cm$^{-2}$ we have determined an unabsorbed X-ray
luminosity for the southern nucleus of $\sim 3 \times 10^{40}$ ergs
s$^{-1}$ in the 0.5-10 keV energy range (see Table 3). This is at
least 2 orders of magnitude below the luminosities of `classical'
Seyfert nuclei (see for example George etal 1998). To suppress the
emission from such an active nucleus requires a much higher column
than the one derived from either the {\it Chandra\/} or the {\it
NICMOS\/} observations.

Alternatively, the observed X-rays could correspond to scattered
emission from a Compton-thick Seyfert 2 nucleus. In this case the
direct emission from the primary source is suppressed by an extremely
large intervening column ($N_{H} > 10^{24}$ cm$^{-2}$) which is
optically thick due to photoelectric absorption (below $\sim 10$ keV)
and Compton scattered above $\sim 10$ keV. In these objects the
central source can only be seen as a reflection component which,
depending on the geometry and the physical conditions around the
nucleus, can originate in either the torus (which is also probably
responsible for the obscuration of the primary source) or a
surrounding ionized medium.

However, one of the strongest indications of reprocessed X-ray
emission in Compton-thick Seyferts is the presence of an iron line
with an equivalent width $> 1$ keV at energies between 6 and 7 keV,
depending on the ionization state of the scattering material (Matt
etal, 2000; Maiolino etal, 1998). The large equivalent width is a
consequence of the continuum being largely suppressed. We have
examined the {\it Chandra\/} data around the position of the southern
nucleus and found {\it no\/} counts in the 6 to 7 keV energy band,
making the scattering scenario unlikely. Finally, recent
high-resolution N-band observations of NGC3256 (Lira \& Ward 2002)
show a weak source ($\sim 50$ mJy) coincident with the southern
nucleus. Its low luminosity is a strong argument against reprocessing
and the presence of a dust enshrouded AGN.

\subsubsection{A low luminosity AGN?}

The presence of a low luminosity AGN (LLAGN) similar to the one
observed in NGC3031 ($L_{\rm x} \sim 2 \times 10^{40}$ ergs s$^{-1}$
in the 2-10 keV band, Ishisaki etal 1996) is still plausible, although
this possibility would not have a significant impact on the overall
energy output of the galaxy. The derived luminosity for the southern
nucleus lies at the lower end of the range of luminosities determined
by Ptak etal (1999) for a sample of LLAGNs and LINERs, and somewhat
larger than the absorption corrected luminosity that Iwasawa etal
(2000) determined for the dwarf Seyfert 1 nucleus in NGC4395. 

To address this scenario we have plotted the extinction corrected
emission from the nuclei of NGC3256 along with the spectral energy
distributions (SEDs) of classical radio loud and radio quite quasars
(from Elvis etal, 1994), LLAGN (adapted from Ho 1999) and starburst
nuclei (from Schmitt etal 1997) - see Figure 9. All the SEDs have been
normalized to the observed nuclear flux in the L band. The nuclear
data correspond to radio observations published by Norris \& Forbes
(1995), JHKL observations reported by Kotilainen etal (1996), HK {\it
NICMOS\/} observations reported in this paper and N-band observations
by Lira \& Ward (2002). Figure 9 shows that both nuclei are too X-ray
weak to be an AGN, while the radio emission has an intermediate value
between the radio loud AGN and LLAGN, and the radio quiet AGN. The
southern nucleus, however, shows a flat distribution between 1 and 10
$\mu$m in $\nu F_{\nu}$ space, which is characteristic of AGN. Given
the large uncertainties in the determination of the X-ray luminosity
for this nucleus, the presence of a LLAGN is certainly possible. The
northern nucleus, on the other hand, presents a sharp rise towards
longer wavelengths, in good agreement with the spectral distribution
for starburst galaxies.

For a brief discussion of the possibility that Inverse Compton
scattering plays a significant role in the X-ray emission from the
nuclei, see section 6.3.4.

\subsection{The discrete source population}

\begin{figure}
\centering
\caption{X-ray isocontours overlaid onto optical HST PC2 images of the
central region in NGC3256.  A broad band F814W ($\sim I$) image is
showed in the top panel and a colour F450W-F814W ($\sim B-I$) image is
showed at the bottom (darker corresponds to redder colours - for a
colour version see Figure 2 in Zepf etal 1999). Both images have the
same size and (non-standard) orientation. Compact sources are labelled
and the position of the southern nucleus is marked with an asterisk in
the top image.}
\end{figure}

\begin{figure}
\centering
\includegraphics[angle=270,scale=0.65]{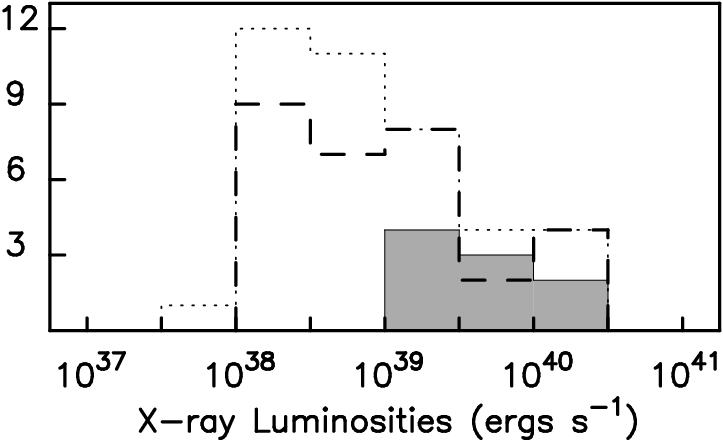}\\ 
\caption{Luminosity distribution of the point source population 
in NGC3256 (solid colour) compared with the observed (thin dotted
line) and binned (thick dashed-dotted line) population seen in the
Antennae galaxy (Fabbiano, Zezas \& Murray, etal 2001). See text for
details.}
\end{figure}

We have used archival HST observations of NGC3256 in order to search
for optical counterparts of the X-ray sources seen in Figure 1. The
galaxy was imaged with the WFPC2 camera on board HST on the 20 of May
1994 and the 6 of September 1994, using the F450W ($\sim B$) and F814W
($\sim I$) broad band filters. The data have already been reported by
Zepf etal (1999).

Figure 10 shows 0.3-10 keV X-ray iso-contours overlaid onto the HST
F814W observations and an F450W-F814W colour map. The {\it Chandra\/}
data have been smoothed using an adaptive smoothing technique
developed by Ebeling, White and Rangarajan (1999) and implemented in
the CIAO software. The algorithm uses a circular Gaussian with
variable width which is adjusted in order to preserve the
signal-to-noise ratio under the kernel.

Several coincidences between X-ray sources and optical knots of
emission can be seen in Figure 10. In the nuclear region tentative
counterparts can be found for most of the X-ray sources, and in
particular for sources 5, 7(N), 9, 10 and 11, which correspond to
moderately absorbed sources, as can be seen in Table 3. On the other
hand, those sources with significantly large values of $N_{H}$
(sources 2, 3 and 8) have no clear optical counterparts. The location
of these absorbed sources in the F450W-F814W colour image shows a
clear correlation between large X-ray hydrogen columns and the
presence of prominent dust lanes.

Photometric measurements for the optical counterparts to the less
obscured X-ray sources show that they are all characterized by blue
colours, with F450W-F814W $\sim -0.7$ to $-1.4$ mags. Using the Space
Telescope Data Analysis System (STSDAS) Synphot task we have also
computed the F450W-F814W colour of the Orion star forming nebula and
found that it is necessary to introduce a visual extinction of $\sim
2.5$ magnitudes to its spectral energy distribution in order to
reproduce the range of colours of the star forming regions in
NGC3256. From the study of the colour-colour diagram presented in
section 3.2, we have derived typical absorption columns towards these
sources of a few times $10^{21}$ cm$^{-2}$.  Assuming a Galactic
gas-to-dust ratio this absorption corresponds to a visual extinction
of $\sim 1-3$ magnitudes.

L\'{\i}pari etal (2000) have obtained optical spectra for several
sources in the nuclear region of NGC3256. Their Region 1 corresponds
to the northern nucleus, while Regions 2, 4, 5, 8 and 9 are probably
the counterparts of the X-ray sources 10, 11, 13, 9 and 6 respectively
(from comparison of their Figure 1c with our H$\alpha$ maps discussed
in section 6.3.1). They found that all spectra are typical of high
metallicity HII regions or starbursts with an extinction of $A_{V}$
$\ga 2$, in good agreement with our results from the {\it Chandra\/}
observations and the HST photometry.

We have found a clear correlation between powerful and compact X-ray
sources and regions of vigorous star formation. A few of the sources
are clearly resolved in the {\it Chandra\/} data and probably
correspond to complex regions of multiple individual sources and
diffuse emission similar to a down-scaled version of the nuclear
region of M82.

Most of the discrete sources seen in the {\it Chandra\/} data are
unresolved. The {\it Chandra\/} spatial resolution (FWHM $\sim 0.5$
arcsec) implies that these point-like sources have linear sizes of
$\la 140$ pc, only a factor of two smaller than the complex X-ray
emitting region seen in 30 Dor, the giant HII nebula in the LMC (Wang
1999). Therefore groupings of less luminous objects contributing to
many of these unresolved sources is likely.

Recent work by Helfand \& Moran (2001), shows that the hard emission
in NGC3256 can be explained if luminous high mass X-ray binaries, like
the ones observed in the Magellanic Clouds, are representative of the
source population. The 2-10 keV X-ray luminosity from all high mass
X-ray binaries in the LMC or SMC is however, only $\sim 3.5 \times
10^{38}$ ergs s$^{-1}$ (Helfand \& Moran 2001), at least one order of
magnitude below the average luminosity of the point sources seen in
NGC3256 (for $\Gamma = 2$, approximately half of the unabsorbed
luminosity will emerge in the 2-10 keV energy range - see Table
3). This implies that at least 10, and up to 30 high mass X-ray
binaries ($L_{\rm x} \ga \times 10^{38}$ ergs s$^{-1}$) need to be
clustered in a region half the size of 30 Dor. This might suggest the
presence of intrinsically luminous X-ray sources in some of the
discrete sources observed in NGC3256, which would dominate their X-ray
output, as is observed, for example, in the M82 starburst galaxy
(Griffiths etal 2000, Kaaret etal 2001, Ward etal 2001).

The presence of these point-like and highly luminous sources in
selected nearby galaxies has been recognised for at least a decade
(Fabbiano 1989), although only recently has the full scope of this
phenomenon emerged from detailed galaxy surveys using high-resolution
X-ray observations (Roberts \& Warwick 2000; Lira, Lawrence \&
Johnson, 2000; Colbert \& Mushotzky, 1999). The nature of these so
called Super Eddington sources (with $L_{\rm x}$ well above the
Eddington limit for a $\sim 1 M_{\sun}$ accreting neutron star) is
still poorly understood. Candidates to explain them include young
supernova remnants evolving in high dense environments, accretion onto
intermediate-mass black holes, mildly beamed (but otherwise normal)
X-ray binaries, `hypernova remnants' (ie the remnants of a Gamma-ray
burst), and compact complexes of several less luminous X-ray
sources. It is quite possible that the Super Eddington sources cannot
be grouped into a single type of object but represent an heterogeneous
class.

{\it Chandra\/} observations of the archetypical starburst galaxy M82
have revealed a highly variable and extremely luminous source ($1 - 10
\times 10^{40}$ ergs s$^{-1}$) which could imply the presence of a $\ga
500$ M$_{\sun}$ compact accreting object (Kaaret etal 2001). Similar
massive accreting black holes could explain some of the Super
Eddington sources found in NGC3256. Unfortunately, the lack of
significant time variability in the compact source light curves
(Figure 5) means that we cannot constrain the luminosity of individual
sources. However, no very short time scale variability has been
detected in the luminous source in M82 suggesting that the dramatic
changes in flux are not flares of short duration (Kaaret etal 2001,
Ward etal 2001). The same situation is found for the Super Eddington
sources detected in the Antennae galaxy (Zezas etal 2001).

\begin{table}
\centering
\caption{Properties of the starbursts in NGC3256 and the Antennae
galaxy. References: (1) Hummel \& van der Hulst 1986, (2) Stanford
etal 1990, (3) Neff \& Ulvestad 2000, (4) Norris \& Forbes 1995, (5)
Heckman, Armus \& Miley 1990. IR luminosities were obtained using IRAS
fluxes (eg, Helou etal 1988) and a distance to the Antennae of 29
Mpc.}
\begin{tabular}{lcccc} \hline
        	& $L_{\rm 6cm}$   		& $L_{\rm IR}$  		& SFR			&Refs\\ 
		& (ergs s$^{-1}$) 		& ($L_{\sun}$)  		& ($M_{\sun}$ yr$^{-1}$)&\\ \hline
Antennae	& $\sim 1 \times 10^{39}$	& $\sim 6 \times 10^{10}$	& 5-6			&1,2,3\\
NGC3256		& $\sim 5 \times 10^{39}$	& $\sim 6 \times 10^{11}$	& 40-80			&4,5\\  
\hline
\end{tabular}
\end{table}

The distribution of observed luminosities of the NGC3256 point source
population is presented as a histogram in Figure 11 (ie, after
discarding the extended sources 2, 9, 10 and both nuclei). Data from
{\it Chandra\/} observations of the Antennae galaxy have also been
included (Zezas etal 2001, Fabbiano etal 2001). The Antennae is the
nearest example of a major merger and shows signatures of powerful
star formation activity in both nuclei and throughout the colliding
disks. The emission from the galaxy is dominated by the starbursts,
although not as powerful as that in NGC3256 (see Table 8).

A total of 42 point sources have been detected in the Antennae and
their observed luminosity distribution is plotted with a thin dotted
line in Figure 11. The observations of this system have a much higher
sensitivity threshold than the observations of NGC3256 and the
population of point sources has been detected down to luminosities
$\sim 10^{37}$ ergs s$^{-1}$. The central regions of this merger are
affected by significant reddening (Mengel etal 2001) and the
distribution of absorbing columns affecting the sources seen in the
Antennae is similar to that found for NGC3256 (Zezas etal 2001),
making the comparison between both galaxies clearly appropriate.

The distance to the Antennae is 29 Mpc - about half the distance to
NGC3256 - and so the different physical sizes of the X-ray emitting
regions probed in both galaxies can introduce significant differences
in the inferred luminosity distribution of the source population.
Hence, to perform a meaningful comparison of the sources seen in both
galaxies we have degraded the resolution of the {\it Chandra\/}
observations of the Antennae to that of NGC3256. The results are
plotted in Figure 11 with a thick dash-dotted line. Only 28 point
sources are found with luminosities between $\sim 10^{38} - 10^{40}$
ergs s$^{-1}$.

The luminosity distribution of sources seen in the Antennae and
NGC3256 are very similar for luminosities above $\sim 5 \times
10^{39}$ ergs s$^{-1}$, but there is a clear deficiency of sources in
NGC3256 below that luminosity, probably due to the lower sensitivity
of the observations. The difference between the unbinned and binned
data for the Antennae suggests that some of the very luminous sources
observed in NGC3256 could be composed of several less luminous
components, as already suggested. This redistribution of sources would
significantly populate the left hand side of the histogram and also
modify its high luminosity end. Nevertheless, the existence of Super
Eddington sources remains a strong possibility for NGC3256.

\subsection{The origin of the X-ray emission}

\subsubsection{Correlation with H$\alpha$ emission}

\begin{figure*}
\centering
\caption{Diffuse (left) and discrete source population (right) X-ray contours 
in the 0.3-10 keV energy range overlaid onto an archive H$\alpha$
image of NGC3256.}
\end{figure*}

In order to further investigate the morphology of the diffuse
component in NGC3256 we subtracted circular regions with a radius of 3
arcsec (which encircles $\sim 90\%$ of the incident photons) on a full
band (0.3-10 keV) image of the galaxy at the positions of the detected
discrete sources to remove their emission. We then interpolated over the
holes and smoothed the image with a Gaussian with $\sigma = 1$ arcsec,
obtaining a diffuse component map. Subtracting this frame of the
diffuse emission from the original image we obtained an image of the
population of compact sources only. This compact source map was then
slightly smoothed ($\sigma = 1$ arcsec).

The two X-ray images were then compared with an archive H$\alpha$
observation of the galaxy obtained with the New Technology Telescope
at La Silla in April 1993. The H$\alpha$ and associated calibration
frames were reduced in the standard manner. The final H$\alpha$ image
was obtained by scaling the `line' and `continuum' images until a good
subtraction of the field stars was achieved.

Figure 12 shows the X-ray contours of the diffuse and discrete source
population overlaid onto the H$\alpha$ image of NGC3256 (the
`roundness' of the contours representing each compact source is an
artifact resulting from the smoothing of the image using a Gaussian
profile). In general, there is a good agreement between the morphology
seen in both wave-bands. The most prominent regions seen in H$\alpha$
are nearly always coincident with enhanced regions of diffuse X-ray
emission. Also, several X-ray compact sources are coincident with
strong H$\alpha$ knots. One interesting feature corresponds to the
H$\alpha$ emission coincident with source 10, where a bubble like
morphology can be seen. On the other hand, no strong H$\alpha$
emission is detected at the position of the southern `Arm' seen in 
diffuse X-rays. Also, none of the compact sources located in this
region (sources 2, 3 and 8) have obvious counterparts in the H$\alpha$
image. As mentioned in section 6.1, $H$, $K$ and Pa$\alpha$
observations of NGC3256 were obtained with {\it NICMOS\/}. The small
field of view of the NIC2 camera was centered around the most obscured
part of the galaxy and contained sources 7, 8, 2 and 3. Again, no
counterparts were found for sources 2 and 3 in any of the {\it
NICMOS\/} images, ruling out a direct association with powerful HII
regions or a young dusty stellar cluster.

A crude flux calibration of the H$\alpha$ image can be achieved using
the long slit observations reported by L\'{\i}pari etal (2000). They
obtained H$\alpha$ fluxes for the brightest knots of emission within
the central region of the galaxy through a slit width of $1.5 - 2.0$
arcsecs. Inspection of the H$\alpha$ image shows that these apertures
contain the bulk of the emission for those knots used in the
calibration.  The spectrophotometric fluxes were compared with the
background subtracted counts obtained from the H$\alpha$ image using a
2 arcsec circular aperture, giving consistent flux-to-count ratios
within a factor of two. Using a large circular aperture ($\sim 41$
arcsec in diameter) we find that the total H$\alpha$ flux detected in
the NTT image corresponds to $\sim 2 \times 10^{-11}$ ergs s$^{-1}$
cm$^{-2}$, or $\sim 7.5 \times 10^{42}$ ergs s$^{-1}$. The H$\alpha$
luminosity is therefore $\sim 300$ times less than the IR luminosity
of the galaxy. Moorwood \& Oliva (1994) estimated that
$L_{IR}/$H$_{\alpha} \sim 175$ after correcting for absorption, and
showed that this value agrees well with models of star formation in
galaxies.

The H$\alpha$ morphology, particularly in the less reddened regions of
the galaxy, is a good indicator of the most active star forming
regions. In particular the good agreement seen between several compact
sources detected both in X-rays and H$\alpha$ is a good indicator of
the presence of young supernova remnants or rich associations of
massive OB stars, and therefore high-mass X-ray binaries.

\subsubsection{The thermal components}

At least two thermal components are necessary to fit the Chandra
spectrum of the diffuse emission in NGC3256. The thermal energy
associated with a thermal plasma can be expressed as $E_{\rm x} =
3\,n_{e} V\,f\,kT$. The electron density, $n_{e}$, can be
derived from the emission integral ($EI$), which is given by the
component normalization during the fit to the X-ray data. Assuming a
purely ionised hydrogen gas, $EI = f\,n_{e}^2V$, where $f$ is the
filling factor, or fraction of the volume $V$ occupied by the X-ray
emitting gas. If this emission is indeed powered by the starburst
activity in the galaxy then $E_{\rm x}$ should be no larger than the
amount of energy that the starburst has injected into the interstellar
medium. This comparison assumes that the injected mass and energy, in
the form of stellar winds and SN ejecta, has been efficiently
thermalised to a characteristic temperature $T$, and evolved into a
galactic wind.

MLH have already shown that the thermal energy stored in the X-ray
emitting gas in NGC3256 is consistent with the amount of energy
injected by the starburst in the last $\sim 10^{7}$ yr ($\sim$ few
$10^{57}$ ergs). The spectral parameters derived from our Chandra
observations for the two thermal components are different from those
derived from the {\it ASCA\/} data (in terms of best fitted
temperatures and normalizations - see las entry in Table 4), but these
differences would only amount to an increase of $\sim 30\%$ in our
estimate of the total $E_{\rm x}$ (by considering both the cool and
warm components). This relatively small change is due to the opposite
effects introduced by the higher temperature we find for the cooler
component (0.6 keV instead of the 0.3 keV derived from the {\it
ASCA\/} data) and the lower normalization we determine for the warm
component ($\sim 2.7$ times smaller than the one derived by
MLH). Still, a larger difference could be introduced if different
geometries are assumed for the regions occupied by the plasma.

In principle, from the spatially resolved spectroscopy, we could
constrain the characteristic sizes of the regions occupied by the
thermal emission, in particular the volume $V$ occupied by the warm
component which was unconstrained by the {\it ASCA\/} observations
(MLH assumed a spherical geometry with $r = 1.4$ kpc centered around
the northern nucleus). Our tentative results show that the 0.6 keV
component dominates for $r \ga 4$ kpc whilst the 0.9 keV gas dominates
over $r \la 3$ kpc, with an intermediate $\sim 0.8$ keV temperature
found in between. The results are less certain if more realistic
errors are computed using 2 parameters of interest for a 90\% CL (ie,
$\Delta \chi^{2} = 4.6$): while region A is characterised by $N_{H}
\sim 0.4 - 1.0$ cm$^{-2}$ and $T \sim 0.6 - 0.9$ keV, region B is
described by $N_{H} \sim 0.1 - 0.5$ cm$^{-2}$ and $T \sim 0.5 - 0.7$
keV. As noted in section 4.2, the result on the $N_{H}$ variation
between the two regions is better constrained than the temperature
variation.

Although not well constrained, our results suggest that the
characteristic radius adopted by MLH for the spherical region occupied
by the warm component could have been underestimated by about a factor
two. The larger volume implies that the thermal energy stored in
the 0.9 keV plasma could be as high as $\sim 1 \times 10^{57}$ ergs
(for a fitted value of $EI = 1.9 \times 10^{64}$ cm$^{-3}$ and $f =
1$). Also, the {\it Chandra\/} data have shown that the cool component
extends further than the 6 kpc radius adopted by MLH, especially
towards the north-east of the galaxy (see Figure 8). The energetics in
this component will be at most $\sim 1.5 \times 10^{57}$ ergs (for $r =
8$ kpc, a fitted value of $EI = 4.2 \times 10^{63}$ cm$^{-3}$ and $f =
1$).

The total energy stored in the thermal components is therefore in good
agreement with the derived energy deposited by the starburst activity
over the last $\sim 10^{7}$ yr. It should be remembered, however, that
not all the kinetic output, in the form of star winds and supernova
ejecta, is effectively thermalised in the form of a galactic
wind. Although difficult to estimate, numerical simulations predict
thermalisation efficiencies of $\ga 50\%$ (Strickland 2001). If this
figure is applicable to NGC3256, then only half of the energy released
by the starburst ($\ga 1 \times 10^{57}$ ergs) would be in the form of
the hot plasma observed with {\it Chandra\/}. However, taking into
account that a volume filling factor close to unity for both thermal
components is highly unlikely, it is still energetically possible for
the starburst to power the observed X-ray emission in NGC3256.

\subsubsection{The galactic wind}

Although the gross properties of galactic winds are well described by
theoretical models, important aspects of the nature of the X-ray
emission are not understood. In particular the detailed origin of the
X-rays is still an open question. In the scenario proposed by
`mass-loading' models the bulk of the X-rays are produced by a wind
which has significantly increased its density and decreased its
temperature due to heavy and efficient mixing with denser interstellar
material (eg, Suchkov etal 1996). In this scenario, the physical
parameters derived for the X-ray emitting material correspond to those
of the wind itself. Support for the `mass-loading' scenario in many
galaxies comes from the comparison between the mass deposition by the
starburst and the X-ray emitting mass (eg, M82 and NGC253 (Ptak etal
1997), Arp220 (Heckman etal 1996), NGC4449 (della Ceca, Griffiths \&
Heckman 1997), Arp299 (Heckman etal 1999)). HLM have shown that the
wind must also be heavily mass-loaded in NGC3256. On the other hand,
some models postulate that the X-ray emission comes from the
interaction of the hot and low density wind with clouds of gas, either
ambient material that has been engulfed within the wind or that is
located at the boundary layer between the flow and the surrounding
medium (eg, Chevalier \& Clegg 1985). In this case the tenuous wind
remains largely undetectable while the X-ray emission is produced by
the low volume filling factor gas from the shocked clouds. If this is
the case for NGC3256, the calculations above should use $f \ll 1$ and
take into account that a large fraction of the wind remains invisible
in our observations. Recent {\it Chandra\/} observations of a
limb-brightened outflow in NGC253 are in good agreement with the idea
that the X-rays are produced in the regions of interaction between the
hot wind and the denser interstellar medium (Strickland etal
2000). {\it XMM-Newton} observations, on the other hand, show that at
lower energies the emission is more uniformly distributed, implying a
higher filling factor and, probably, a `mass-loaded' wind (Pietsch
etal 2001).

\subsubsection{The temperature gradient}

Finally, our results in section 4.2 have suggested a gradient of
temperature with radius. A similar distribution of temperatures has
already been found in a study of the starburst galaxy M82 by
Strickland, Ponman \& Stevens (1997). They determined a temperature
profile given by $T \propto z^{-\alpha}$, with $z$ the distance along
the minor axis of the galaxy, and $\alpha \sim 0.2 - 0.5$. Given the
poorly constrained temperatures in NGC3256 only a rough profile can be
derived. Adopting a characteristic temperature of 0.9 keV at the
projected distance $r \sim 1$ kpc and 0.6 keV at $r \sim 6$ kpc gives
a power law index $\alpha \ga 0.2$.  We cannot compare this value with
the ones derived for M82 or with theoretical predictions until we take
into account the projection effects, ie, the superposition of
different regions of the flow in our two-dimensional view of the
starburst. However, the nearly spherical geometry of the wind coupled
with an axis of the outflow almost coincident with the line of sight,
as derived from kinematic studies (L\'{\i}pari etal 2000), implies
that a flux-weighted average of the temperature along the galaxy minor
axis returns exactly the same power law index as the original
spherical distribution: assuming that the intensity of the emission is
proportional to the density squared ($I \propto \rho^{2}$) and that
$\rho \propto R^{-\beta}$ and $T \propto R^{-\alpha}$, with $R$ the
radial distance from the centre of the starburst in spherical
coordinates, it follows that

\[ \frac{\int_{z=0}^{\infty} IT dz}{\int_{z=0}^{\infty} I dz}\propto
\frac{1 / r^{2\beta + \alpha -1}}{1 / r^{2\beta -1}} \propto r^{-\alpha}
\]

with $R^{2} = r^{2} + z^{2}$.  In the case of M82, which is viewed
nearly edge on, Strickland, Ponman \& Stevens (1997) adopted a
cylindrical geometry for the wind but showed that if the flow
properties follow a conical or spherical geometry instead then the
measured temperature profile will be slightly flatter than the real
distribution. Our result for NGC3256 is within the range of values
found for M82 and is much flatter than the prediction of $\alpha =
4/3$ using adiabatic models (Chevalier \& Clegg 1985). However, as
explained above, the Chevalier \& Clegg model postulates that the
X-ray emission does not come from the wind itself, but from shocked
clouds of gas. The temperature reached by these clouds will depend on
the speed of the shock driven into them by the wind. A shock heated
cloud model is indeed in good agreement with the results obtained for
M82 (for further discussion see Strickland, Ponman \& Stevens 1997).

\subsubsection{The hard emission}

MLH have argued that the 2-10 keV intrinsic luminosity in the power
law component ($\sim 2 \times 10^{41}$ ergs s$^{-1}$ as measured by
{\it ASCA\/}) could not be explained solely by the presence of a
Galactic population of high mass X-ray binaries and that Inverse
Compton scattering was likely to make a significant contribution. They
reached this conclusion by comparing the hard X-ray luminosity per O
star observed in our Galaxy with the one inferred for NGC3256. New
results published by Helfand \& Moran (2001) show that in fact the
2-10 keV luminosity is well explained if a Magellanic Cloud type
population of high mass X-ray binaries is assumed instead of a
Galactic population.

{\it Chandra\/} observations show that most of the power law component
seen with ASCA can be identified with a population of discrete
sources. These extremely bright sources can have luminosities at least
one order of magnitude above any source seen in our Galaxy. From a
comparison of the luminosity in the discrete source population and the
diffuse component in the 2.5-10 keV range we find that $\sim 75-80\%$
of the emission ($\sim 1.2 \times 10^{41}$ ergs s$^{-1}$) comes from
the hard ($\Gamma \sim 2$) and powerful ($10^{39}-10^{40}$ ergs
s$^{-1}$) compact sources resolved by {\it Chandra\/}. A few of these
sources have been found to be extended and therefore they probably
correspond to compact star-forming regions. Since it is becoming
increasingly evident that super Eddington sources are often found in
galaxies undergoing vigorous episodes of star formation, we can
speculate that some of the observed discrete sources found in NGC3256
correspond to (or are dominated by) these high luminosity sources. It
is, however, impossible to test this further with the present data.

The extended emission has a power law component characterised by a
fairly steep slope ($\Gamma \la 3$, or $kT \sim 4$ keV if a Mekal
model is adopted). This component represents $\sim 75\%$ of the total
luminosity in the diffuse emission in the 2.5-10 keV band (see Table
5). An image in this energy range would determine the morphology of
the $\Gamma \sim 3$ emission, but our data do not have sufficient
counts for this. From the spectral fit in Table 5, however, we know
that a moderate extinction is associated with this component, and
therefore it is unlikely that the bulk of the emission is connected
with the innermost, and more heavily reddened parts of the starburst.
Also, the inclusion of this component in the fit to the different
spatial regions seen in Figure 8 always gave acceptable values of
$\chi^{2}_{\rm red}$, suggesting that the emission is present even at
large radii. Therefore, the $\sim 4 \times 10^{40}$ ergs s$^{-1}$
emitted in the 2.5-10 keV range in the power law component is most
likely due to an unresolved population of X-ray binaries and supernova
remnants distributed throughout the galactic disk. The inferred
temperature is at the top end of the range of temperatures measured
for young supernova remnants ($kT \sim 1-4$ keV; Bregman \& Pildis
1992) and is close to the temperatures observed in low mass X-ray
binaries ($kT \sim 3$ keV; Barret 2001). Inclusion of some
contribution from the harder spectra of high-mass X-ray binaries and
X-ray pulsars would improve the spectral match.

\subsubsection{Inverse Compton Scattering?}

Our {\it Chandra\/} observations have shown that most of the $\Gamma
\la 2$ power law emission is accounted for by hard and powerful
discrete sources. So far we have speculated that accretion onto
compact objects is the dominant mechanism responsible for this
emission. However, Inverse Compton (IC) scattering is also a viable
non-thermal mechanism that could contribute significantly.

NGC3256 shows strong IR and nonthermal (Synchrotron) radio emission.
This implies the spatial co-existence of a population of relativistic
electrons and IR seed photons which could be up-scattered by the
electrons into X-ray frequencies. The IC luminosity can be expressed
as $L_{{\rm IC}} = L_{{\rm Syn}} \times U_{{\rm ph}} /\, U_{{\rm B}}$,
where $L_{{\rm Syn}}$ is the radio luminosity and $U_{{\rm ph}}$ and
$U_{{\rm B}}$ are the seed photon and magnetic energy densities,
respectively (cf, Rybicki \& Lightman 1979). The low surface
brightness of the radio and IR emission outside the nuclei implies
that IC emission is probably negligible in the extranuclear regions.
For example, source 10 in Figure 1, the brightest source seen outside
the nuclei at X-ray and optical wavelengths (for optical broad band
data see Table 3 in L\'{\i}pari etal 2000), appears much fainter than
the nuclei in the $L$ band map presented by Kotilainen etal (1996) and
in the radio maps by Norris \& Forbes (1995), implying a small value
for the $L_{{\rm Syn}} \times U_{{\rm ph}}$ product.

In the case of the nuclei, however, IC emission could still play a
significant role in the production of hard X-rays. This is
particularly relevant in the case of the northern nucleus, which has a
2-10 luminosity of $\sim 2.5 \times 10^{40}$ ergs s$^{-1}$ (which is
completely dominated by the power law component) and has been found to
be an extended source in X-rays (FWHM $\sim 1$ arcsec). A similar size
has also been determined from radio and N-band (12 $\mu$m)
observations (Norris \& Forbes 1995, Lira \& Ward 2002), although a
much smaller peak is determined from NICMOS data at 2.2 $\mu$m
(section 6.1.1). MLH have shown that between $5 \times 10^{39}$ and $5
\times 10^{40}$ ergs s$^{-1}$ could be produced by IC scattering in
each nucleus in the 2-10 keV band.

If IC scattering does take place, the power law spectra of the radio
and X-ray emission are expected to have the same slope, since the
index of the power law distribution of the electron energies ($n$)
determines the output spectrum for Synchrotron and IC radiation in the
same manner ($\alpha = (n-1)/2$, where $\alpha$ is the power-law
energy index of the out-coming spectra). From the radio observations
of NGC3256 (Norris \& Forbes 1995) we find that the power-law index of
the radio emission in the nuclei has $\alpha \sim 0.8$. This value is
inconsistent with the X-ray index observed in the northern nucleus
power law ($\alpha = \Gamma - 1 \sim 1.2 - 2.4$ - see Table 2). More
realistic errors, with 3 parameters of interest for a 90\% CL (ie,
$\Delta \chi^{2} = 6.3$) imply that $\alpha \sim 1 - 2.8$, with a
lower limit closer to the value of the radio index. No detailed
comparison can be made for the southern nucleus, although the (poorly
constrained) range of power-law indices derived in section 3.2 is
consistent with the observed radio slope. Therefore, IC emission
cannot be dismissed as a possibly significant component of the
observed hard luminosities from both nuclei, although this emission
can also be explained solely by the output from accretion driven
sources, and in particular, by the presence of super-Eddington
sources.

\section{Conclusions and summary}

Until recently, the detailed study of the diffuse emission and point
source population in starburst galaxies has been a difficult task. The
poor spatial resolution and limited wavelength coverage of X-ray
missions prior to {\it Chandra\/} have resulted in an incomplete
picture. Einstein and ROSAT images were able to detect a diffuse
component and resolve some of the point sources in several starburst
galaxies, but their narrow spectral energy range gave only partial
information on the complex multi-components making up the
emission. Satellites with wider spectral coverage ({\it ASCA,
BeppoSAX\/}), were able to study the X-ray emission of starburst
galaxies up to harder energies, but their extremely low spatial
resolution gave little or no information about the different spatial
components. This situation is now changing dramatically as {\it
Chandra\/} data for starburst galaxies become available.

In this paper we have presented a detailed spatial and spectral
analysis of {\it Chandra\/} observations of the IR luminous merger
system, NGC3256. Our main results can be summarized as follows:

(1) The total absorption corrected luminosity of the galaxy over the
0.5-10 keV energy range is $L_{\rm x} \sim 9 \times 10^{41}$ ergs
s$^{-1}$ ($\sim 50\%$ of the luminosity inferred from previous {\it
ASCA\/} observations due to the different adopted models). A
significant fraction ($\sim 80\%$) of the X-ray emission is detected
as diffuse emission. Fourteen compact sources are detected, with
absorption corrected luminosities between $\sim 10^{39}$ and $\sim
10^{41}$ ergs s$^{-1}$. The X-ray spectra of the diffuse emission is
predominantly soft, while the discrete sources exhibit
characteristically harder spectra.

(2) Both galaxy nuclei are clearly detected in X-rays. The northern
nucleus is spatially resolved (FWHM $\ga 270$ pc) and it corresponds
to the brightest source, with a corrected luminosity of $\sim 1 \times
10^{41}$ ergs s$^{-1}$ in the 0.5-10 keV range. The southern nucleus
is a heavily obscured source and with an intrinsic luminosity $\sim
3-4 \times 10^{40}$ ergs s$^{-1}$.  No clear evidence is found for the
presence of an AGN in the southern nucleus.

(3) All the detected discrete sources have luminosities one order of
magnitude above the Eddington limit for a $\sim 1 M_{\sun}$ accreting
neutron star. No doubt fainter point sources are present but are below
the sensitivity threshold of our observations. No variability was
detected in any of the sources found in NGC3256 and therefore, we were
unable to constrain the luminosity of {\it individual \/}
sources. However, intrinsically powerful sources ($L_{\rm x} \ga
10^{39}$ ergs s$^{-1}$) are very likely to be present.

(4) The diffuse emission has a soft spectrum which can be described as
the superposition of two thermal plasma components with temperatures
$T \sim 0.6, 0.9$ keV plus a harder tail with $\Gamma \sim
3$. Tentative evidence is found from spatially resolved spectra for a
trend in the extinction and the temperature of the thermal plasma to
decrease with larger radial distances from the nuclear region, in
agreement with observations of other starburst galaxies and with wind
models.

(5) The morphology of the diffuse emission and the location of the
discrete sources show a good correlation with optical and H$\alpha$
images. The extinction determined from the X-rays is in good agreement
with the values derived from optical and infrared NICMOS data.

(6) While the thermal components in the diffuse emission can be
explained in terms of the mass and energy deposition by the nuclear
starburst, the observed hard tail is best explained by a combination
of a population of faint point sources and hot plasma emission from
young SNRs.  Inverse Compton scattering could still be important in
explaining the hard X-rays seen in the two nuclei.

\section*{Acknowledgements}{We thank Pepi Fabbiano for kindly making
available results from {\it Chandra\/} observations on the Antennae
galaxy. The archival data are based on observations made with ESO
Telescopes at the La Silla under programme ID 51.2-0036 and on
observations made with the NASA/ESA Hubble Space Telescope, obtained
from the data archive at the Space Telescope Institute. STScI is
operated by the association of Universities for Research in Astronomy,
Inc. under the NASA contract NAS 5-26555. This work was supported by
CXC grant GO1-2116X. AZ acknowledges support by NASA contract
NAS8-39073 (CXC). Data were reduced using Starlink facilities.}

\bibliographystyle{/home/ltsun0/plt/tex/mnras/mnras}
{}

\end{document}